\documentclass{aa}  

\usepackage{graphicx}
\usepackage{txfonts}
\begin{document}

   \title{Irradiation induced mineral changes of NWA10580 meteorite determined by infrared analysis}


   \author{I. Gyollai
          \inst{1,4}
           \and
          S. Biri\inst{2}
          \and
          Z. Juh\'asz\inst{2}  
          \and
          Cs. Kir\'aly \inst{4,6}
          \and
          B. D. P\'al \inst{7,3,4}       
          \and
          R. R\'acz \inst{8}
          \and
          D. Rezes \inst{3,4,5}
           \and
          B. Sulik\inst{2}
          \and
          M. Szab\'o \inst{1,4}
          \and
          Z. Szalai \inst{4,6}
          \and
          P. Sz\'avai \inst{4,6}
           \and 
          T. Szklen\'ar \inst{3,4}  
           \and
          \'A. Kereszturi \inst{3,4}
             \fnmsep\thanks{Corresponding author: kereszturi.akos@csfk.org}
          }

   \institute{Institute for Geological and Geochemical Research, Research Centre for Astronomy and Earth Sciences, HUN-REN, 
              Budapest, Hungary \\
         \and
             HUN-REN, Institute for Nuclear Research            
             Debrecen, Hungary\\
        \and
            Konkoly Thege Astronomical Institute, Research Centre for Astronomy and Earth Sciences, HUN-REN, 
            Budapest, Hungary\\
        \and 
            CSFK, MTA Centre of Excellence
            Budapest, Konkoly Thege Miklós út 15-17., H-1121, Hungary\\
        \and
            Eotvos Lorand University of Sciences (ELTE)
            Budapest, Hungary\\
        \and    
            Geographical Institute, Research Centre for Astronomy and Earth Sciences, HUN-REN, 
            Budapest, Hungary\\           
        \and
            Department of Systems Innovation, School of Engineering, The University of Tokyo, Japan\\
        \and    
            Atomki, Institute of Nuclear Research, 4026, Debrecen, Hungary\\
            }

   \date{Received September 15, 1996; accepted March 16, 1997}

 
  \abstract
   {\\Identifying minerals on asteroid surfaces is difficult as space weathering modifies the minerals' infrared spectra. This should be better understood for proper interpretation.}
   {We simulated the space weathering effects on a meteorite and recorded the alterations of the crystalline structure, such as the change in peak positions and full width at half maximum values.}
   {We used proton irradiation to simulate the effects of solar wind on a sample of NWA 10580 CO3 chondrite meteorites. After irradiation in three gradually increased steps with 1 keV ion energy, we used infrared microscopic reflectance and diffuse reflectance infrared Fourier transform spectroscopy (DRIFTS)  to identify and understand the consequences of irradiation.}
   {We find negative peak shifts after the first and second irradiations at pyroxene and feldspar minerals, similarly to the literature, and this shift was attributed to Mg loss. However, after the third irradiation a positive change in  values in wavenumber emerged for silicates, which could come from the distortion of SiO$_4$ tetrahedra, resembling shock deformation. The full width at half maximum values of major bands show a positive (increasing) trend after irradiations in the case of feldspars, using IR reflection measurements. Comparing DRIFTS and reflection infrared data, the peak positions of major mineral bands were at similar wavenumbers, but differences can be observed in minor bands .}
   {We  measured the spectral changes of meteorite minerals after high doses of proton irradiation for several minerals. We show the first of these measurements for feldspars;  previous works only presented pyroxene, olivine, and phyllosilicates.}


   \maketitle
%

\section{Introduction}

   Space weathering effects on airless body surfaces include irradiation, implantation, sputtering processes due to galactic and solar energetic particles, UV irradiation, meteorite impacts, and temperature fluctuations. These processes modify the appearance and crystalline structure of minerals as well as the composition of asteroid surfaces due to mechanical fracturing, melting, ablation, ion implantation, and destruction of the crystalline lattice, leading to   amorphization.

  The speed of the solar wind strongly influences the consequences of space weathering; it also influences  the potential energy of the solar wind ions and the total witnessed dose. The fluence values are representative for solar wind exposure  timescales  (e.g., \citealt{Dukes1999,Vernazza2013, Lantz2015, Lazzarin2006, Brunetto2014, Lantz2017}). Since the timescales of laboratory tests are substantially shorter than  typical astrophysical exposure durations, higher doses and more intense radiations are used to gain realistic results. In  laboratory experiments the irradiations can be characterized by various physical quantities. The flux, which is the number of ions per surface area deposited during the unit time, determines the rapidity of the surface processing. The induced change is a function of the total fluence,  which is  the total number of ions per surface area deposited during the irradiation. Alternatively, the total deposited energy per surface area can be used, or the dose, which is the absorbed energy by a unit mass of the irradiated material. The recorded infrared spectral curves before and after irradiation tests indicate modifications caused by mineral changes, which we discuss in this work.

   We summarize earlier   irradiation test experiments below to provide context. Both infrared and Raman spectroscopy techniques provide information on the mineral structures of the targets, usually different sample materials 
were measured (e.g., olivine,  \citet{Lantz2017};  polystyrol and olivine,  \citet{Kanuchova2017}). The OH loss from minerals can be better detected by infrared spectroscopy, and less crystallized minerals (clay minerals, ferrihydrite) have a more extended literature background using infrared spectroscopy. With Fourier transform infrared (FTIR) spectroscopy using the diffuse reflectance infrared Fourier transform spectroscopy (DRIFTS)  method, almost bulk (millimeter- to centimeter-sized) meteorite chips were usually measured \citep{Lantz2017,Brunetto2014, Brunetto2020}. The FTIR spectra of Frontier Mountain 95002 and Lancée meteorites are similar, and they both showed typical olivine bands (830–860 cm $^{-1}$) corresponding to a composition of Fo-50–60 \citep{Hamilton2010} before irradiation, which shifted to Fo-30–35 after irradiation \citep{Brunetto2020}. This was interpreted as Mg loss from olivine after irradiation caused by amorphization of the structure. The doublets observed at about 500 cm $^{-1}$ (20 $\mu$m) and 400 cm $^{-1}$ (25 $\mu$m) in CO/CV and at about 450 cm $^{-1}$ (22.2 $\mu$m) (saponite) in CI/C2 are modified by irradiation \citep{Brunetto2020}. Ion irradiation experiments on Tagish Lake using 1 $\times10^{16}$ He$^+$ cm$^{-2}$ at 200 keV reported \citep{Vernazza2013} Mg-rich phylosilicate band shifts of about 0.3 $\mu$m for the 9.8 $\mu$m (10-20 cm $^{-1}$), and about 0.15 $\mu$m for the 22.3 $\mu$m (448 cm $^{-1}$) band \citep{Vernazza2013}. The main band at about 445–448 cm $^{-1}$ (22.32–22.47 $\mu$m), measured on the Mighei meteorite, shows the largest spectral shift \citep{Brunetto2020} due to irradiation.

   Ion irradiation is known to amorphize silicates, as shown by IR-spectroscopy analysis \citep{Demyk2004}. \citet{Dukes2015} observed an increase in  the Fe/Si ratio on their Tagish Lake sample after ion irradiation. \citet{Lantz2015} found shifts of the phyllosilicate and olivine bands near 2.7–10 $\mu$m toward the Fe-rich spectral region, suggesting a loss of the element Mg, probably due to a preferential replacement by Fe. The spectral shift of the IR bands could   also be due to a loss of magnesium, as a result of preferential sputtering (when the somewhat more volatile Mg is more easily sputtered backward than the heavier Fe) \citep{Hapke1975}, possibly leading to amorphization indicated by the broadening of the bands, seen and confirmed by the Raman method \citep{Brucato2004, Demyk2004}.

   Here our aim is to identify and to understand the consequences of artificial solar wind simulating proton irradiation on a sample of NWA 10580 CO3 chondrite meteorites, and the infrared analysis based mineral changes. Since the cosmic weathering has a substantial influence on the spectra of meteorite host asteroids to a large extent, we focused on artificial irradiation to understand the spectral modifications, and  applied three gradually increased irradiation levels. In addition, the survey of the observability of different bands at different spectral resolutions also provides information on the observational capabilities that are needed for an infrared detector on board any future asteroid mission that aims to identify the consequences of space weathering. Although   several works  on mineral changes caused by artificial irradiation have been published, few of them apply gradually increasing irradiation as we did to follow the alteration process closely. In addition, while in most of the  previous works  bulk infrared analysis after irradiation was performed, in this work separate minerals were surveyed.


\section{Methods}

    We used two basic strategies for this work. The first was to irradiate the meteorite by gradually increased doses of solar wind simulating artificial particle bombardment, and the second was to follow the changes of its infrared spectral characteristics by comparing the measurements both before and after each irradiation action.

    We analyzed a 11x8x2 mm   section of the NWA 10580 CO3 chondrite meteorite in this work. This is a poorly altered primitive meteorite with unweathered material, mainly composed of forsteritic olivine, enstatite, and diopside, with many small chondrules, very few CAIs, sulphide blobs, all embedded in a fine-grained matrix. The surface was rough plain shaped by cutting process but not polished, resulting a moderately flat surface with submillimeter undulations. The same side of this meteorite slice was irradiated and measured subsequently in three steps. The first measurement was made before irradiation to see its intact infrared spectra; after each irradiation the same sequence of measurements was realized with DRIFTS and reflection methods. By using DRIFTS the bulk spectra were recorded. Measurements using the  microscope-based reflection method were realized of an $\sim$0.1$\times$0.1 mm  subsection of the meteorite sample.

\begin{figure}
   \centering
   \includegraphics[width=\hsize]{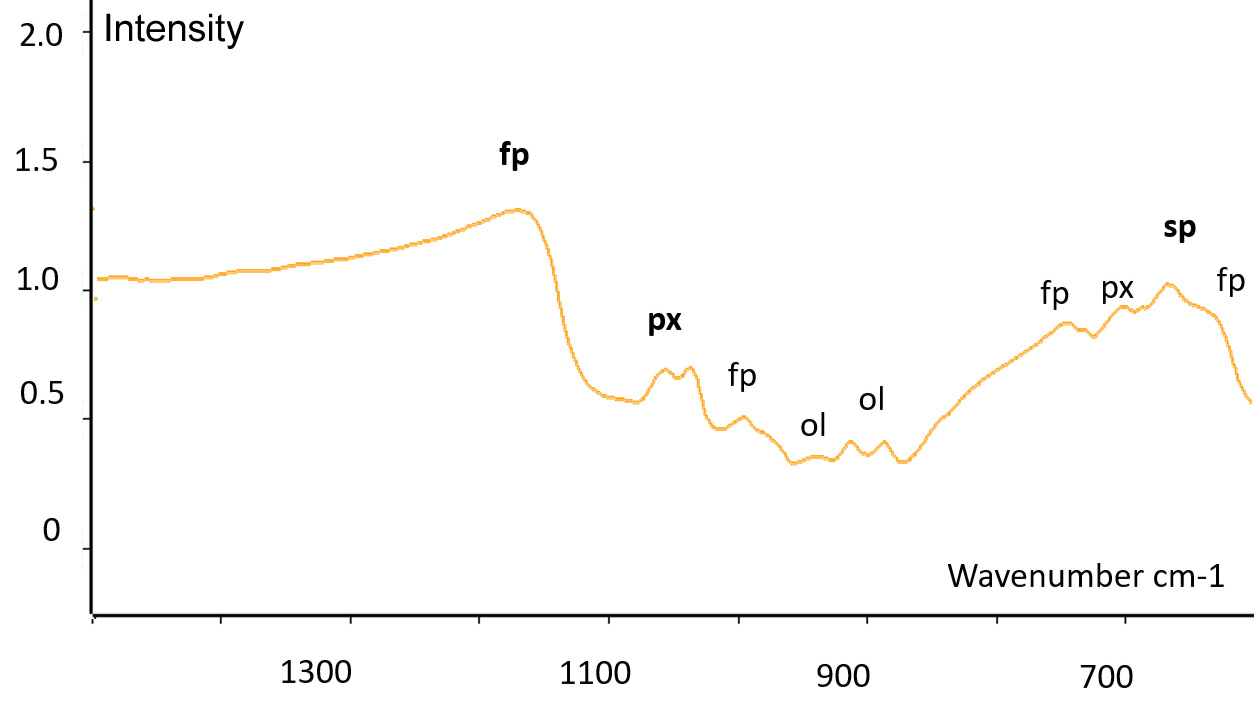}
      \caption{Example spectra of meteorite NWA 10580  with the main peaks (labels in bold) and some of the minor peaks  before irradiation.}
         \label{fig:example}
   \end{figure}

    For infrared spectroscopy and microscopy we used a Vertex 70 FTIR spectrometer and a Hyperion 2000 microscope with 15x IR objective in reflection mode \citep{Biri2012}. For the measurements we did 32 scans in the 400-4000 cm$^{-1}$ range, with 30 s duration at 4 cm$^{-1}$ spectral resolution. We used the  Bruker Optics' Opus 5.5. software to manipulate the  spectra (e.g., baseline correction, atmospheric compensation). The absorbance maximum observed in this work is at the position in wavenumber corresponding to the strongest v1 SiO$_4$ vibration. The areal view of the IR 15x objective is 200 $\mu$m and this method is applicable for polished rock samples. Minerals were identified by the RUFF Database and the Crystal Sleuth software \citep{Lafuente2016}, as well as by the relevant publications (see Fig. \ref{fig:example}. for example peaks).

    For the DRIFTS measurements we also used  the  Vertex 70 FTIR  spectrometer with a Praying Mantis diffuse reflectance accessory. The following parameters were usually used: a spectral resolution of 4 cm$^{-1}$, a sample scan time of 512 s, and a covered wavelength range of 4000–400 cm$^{-1}$. The DRIFTS instrument operates by directing the IR radiation into a sample cup, where it interacts with the sample grains. The IR light is reflected or scattered off as it scans the surfaces, and becomes diffused  \citep{Drocher2012}. The spatial resolution of the measurements was about 1 mm, but definitely below 1.5 mm due to the infrared spot size on the sample. The IR radiation is reflected from the sample surface in all directions, and therefore DRIFTS requires a special mirror arrangement \citep{Mitchell1993,Fraser1990,Korte1988,Kortum1964}. The output mirror directs this scattered radiation to the detector in the spectrometer to record the collected IR light as an interferogram signal, which is used to generate a spectrum. Typically, a background is collected with the empty cup DRIFTS accessory or filled with just potassium bromide (KBr).

    The full width at halm maximum (FWHM) values or the absorption bands were determined by manually measuring the width at half height. The peak shifts were determined in these cases, where the bands appeared before and after irradiations as well (Tables SOM II/1-2, \ref{tab:3}). The irradiation effects of space weathering were simulated by 1 keV protons produced by the electron cyclotron resonance (ECR) ion source at ATOMKI with three different levels of fluence: 10$^{11}$ ion/cm$^2$, 10$^{14}$ ion/cm$^2$, and 10$^{17}$ ion/cm$^2$. The samples were irradiated under vacuum conditions; the pressure was below 1$\times$10$^{-7}$ mbar in the experimental chamber. The first irradiation lasted only for a few seconds; the third lasted for almost one day. In order to perform the different irradiation steps in reasonable time durations, the ion beam current was set at different values from 20 nA to 3 $\mu$A. This corresponds to fluxes ranging from 3$\times$10$^{10}$ to 4.5$\times$10$^{12}$ ions/cm$^2$/s on the irradiated area of 4 cm$^2$.

\section{Results}

    In this section we first present the mineral identifications, then we summarize the changes in peak positions related to the irradiation series, and finally we describe the changes in FWHM values caused by the irradiation. In several cases the measurement series after the irradiation actions did not show a similar and strong trend, but presented a certain variability in the measured quantities. Thus, we focus on the general characteristic changes, which were strong and coherent in the observed cases,  and describe them at the beginning of each subsection. We mention the more stochastic and diverse changes, which are more difficult to interpret, at the end of each subsection. We include the related numerical tables as supporting online materials.

\subsection{Mineral identification}

    The following main mineral bands and FWHM values were used for statistics: 849, 880 cm$^{-1}$ for olivine; 1049 cm$^{-1}$ for pyroxene; 1150 cm$^{-1}$ for feldspar. Certain bands can occur in several minerals:  664 in spinel and pyroxene; 1000 and 980 in olivine and feldspar. The FWHM is also influenced  by the possible occurrence of different mineral phases with nearby or overlapping peak positions. Merged peaks from other mineral phases produce lower FWHM values;  for example, the 670 cm$^{-1}$ band of spinel has a lower FWHM value, due to the occurrence of bands of other minerals such as the 700 cm$^{-1}$ band of pyroxene, and the 635 cm$^{-1}$ and 745 cm$^{-1}$ bands of feldspar. 
    
   \begin{figure}
   \centering
   \includegraphics[width=\hsize]{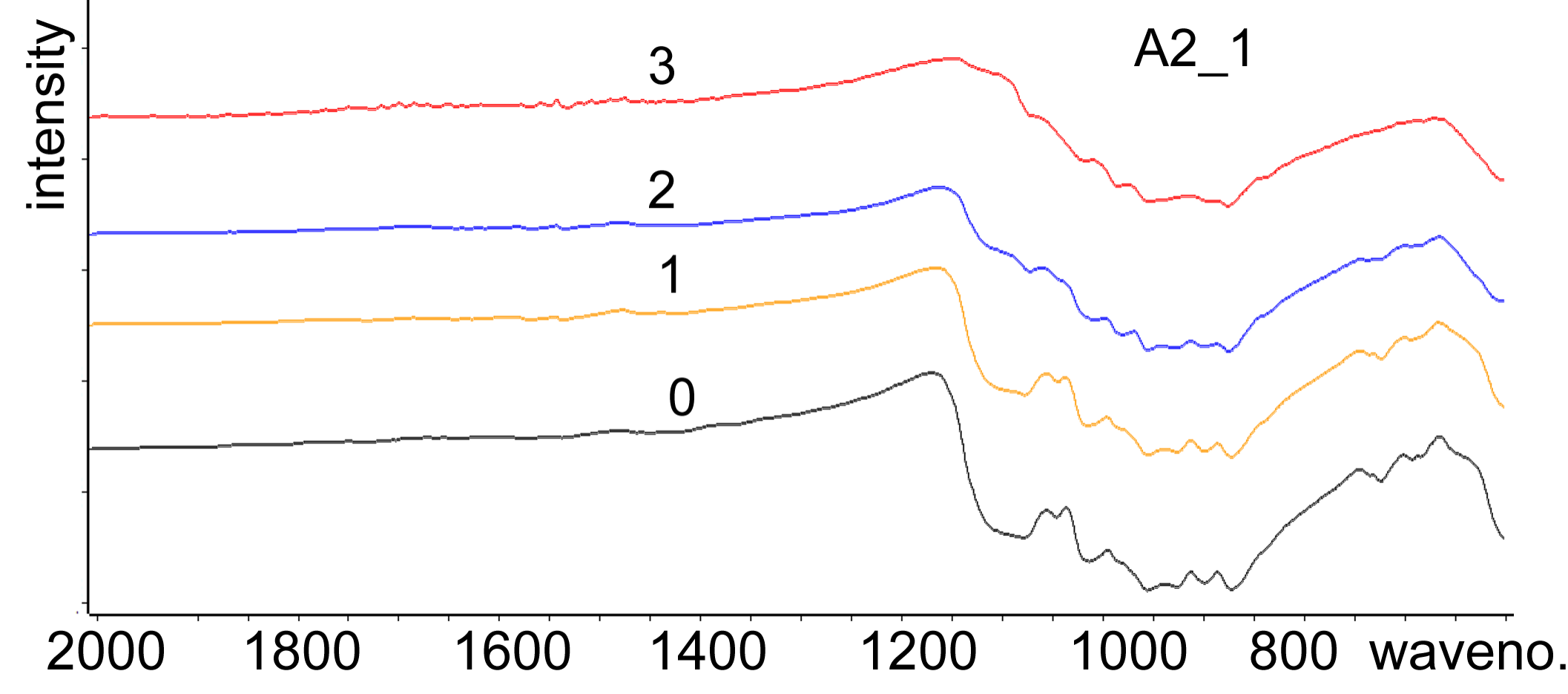}
      \caption{Reflection spectra of the  A2\_1 area:   before irradiation (0), and after the  first (1), second (2), and third (3) irradiations.}
         \label{fig:DRIFT}
   \end{figure}

    Depending on the amorphization level, the minerals may or may not be detected by minor bands in addition to the major bands. For example, olivine might show only one band from a major doublet at a high  amorphization level. In general, the minerals (olivine, pyroxene) had fewer bands after the third (strongest) irradiation than before (Fig. \ref{fig:DRIFT}). The first and second irradiations provided smaller doses and resulted in smaller alterations, and hence the changes in peak positions could be minor and happen along crystal lattice weakness, along with possibly increased noise as the changes are partly stochastic. The consequences of each irradiation action  (from before the first irradiation to after the third) can be seen in the example figures in the subsections below;   in some cases the original trend changed from the early meta-stable state toward the strongest last irradiation action. 

\subsubsection{Olivine}

\begin{figure}[h]
        \centering
        \includegraphics[width=\hsize]{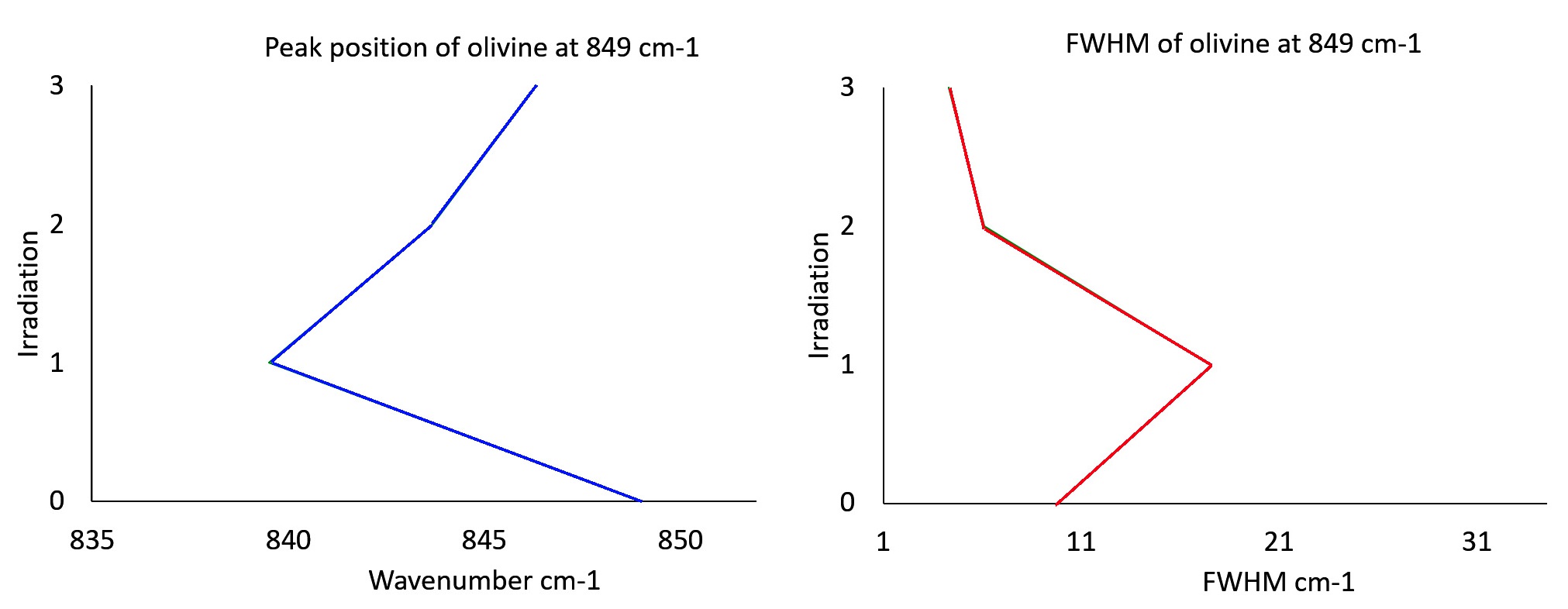}
        \caption{Olivine 849 cm$^{-1}$ peak position changes (blue line, left) and FWHM changes (red line, right), with the number of irradiations on the Y-axis.}
        \label{fig:olivine849}
    \end{figure}

\begin{figure}[h]
        \centering
        \includegraphics[width=\hsize]{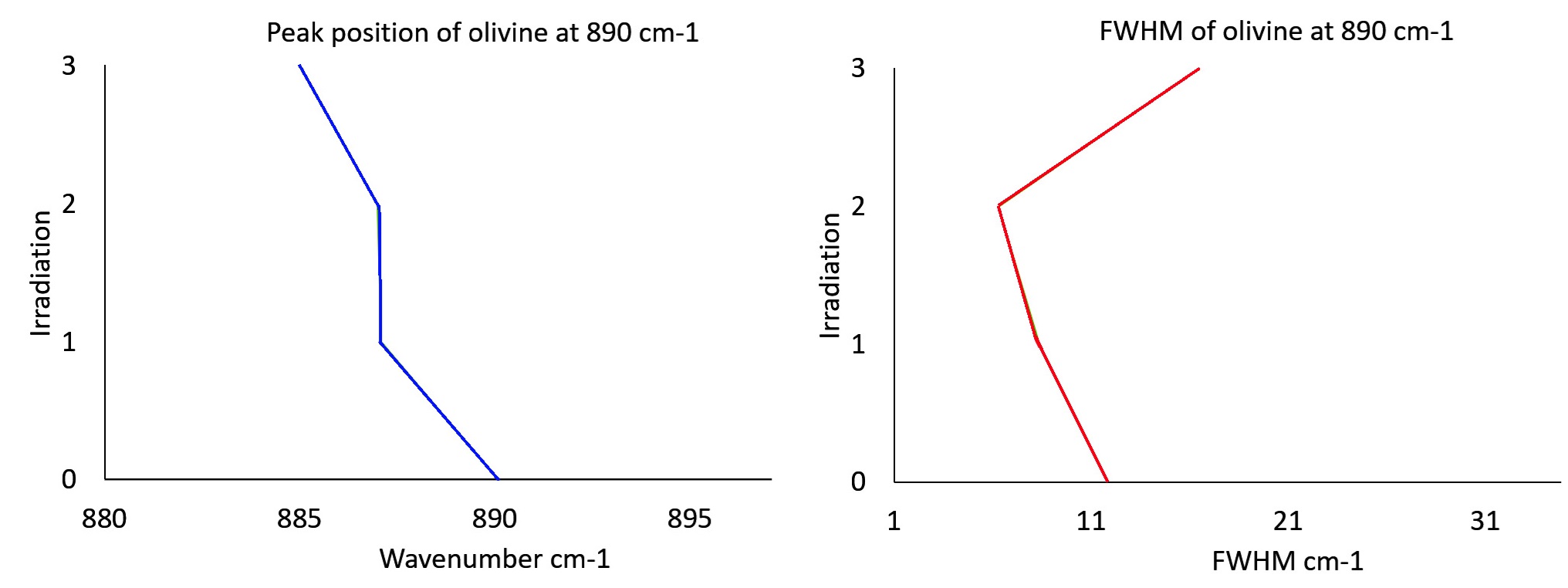}
        \caption{Olivine 890 cm$^{-1}$ peak position changes (blue line, left) and FWHM changes (red line, right), with the number of irradiations on the Y-axis.}
        \label{fig:olivine890}
    \end{figure}

    Before irradiation the bands of olivine appear at 884-894, 913, 940, and  978 cm$^{-1}$. The same bands appear only after the second irradiation (c1 area spectra 1-6). In some cases (at certain locations) all of the bands disappear after the irradiation actions (c1 area spectra 7-9, a2 area spectrum 2), and in other cases the olivine appears only after the irradiation (R1  area spectrum 6, b1  area spectra 4-5, d1 area spectra 5-8) (Table SOM I/5). The 884 cm$^{-1}$ band has a shift of +6 cm$^{-1}$ after the first irradiation (d1/1-4), -3 cm$^{-1}$ after the second irradiation (d1/1-4), and -4 cm$^{-1}$ after the third irradiation (b1/5, d1/1-2) (Tables \ref{tab:1},\ref{tab:2}, Tables SOM I/1-4). In the a2 chondrule the 885 band is detected only after the first irradiation, with a peak shift of -2 cm$^{-1}$; after the second and third irradiations this band disappears (Tables \ref{tab:1},\ref{tab:2}). In the a3 chondrule the 849 band shifts by -15 cm$^{-1}$ after the first irradiation,  increases by 15cm$^{-1}$ after the second irradiation;  the peak shift is -4 cm$^{-1}$ after the third irradiation. To summarize, the peak shift could be observed where the above-described peak positions appear more than once after the irradiation and do not disappear immediately. Some examples of the changes can be seen in Fig. \ref{fig:olivine849} and Fig. \ref{fig:olivine890}. 

    Depending on the structure and composition of olivine, in most cases the major bands (between 894 and 837 cm$^{-1}$) appear. The 890 cm$^{-1}$ band appears before and after irradiation as follows: before irradiation, 14 times; after the  first irradiation, 9 times; after the second irradiation, 6 times; after the third irradiation, 8 times. In general, the appearance of the 890 cm$^{-1}$ band of olivine becomes rarer after the irradiations. The 845 cm$^{-1}$ band shows more stochastic changes; its appearance before and after irradiation shows the following numbers: before irradiation, 6 times; after the first irradiation, 3 times; after the second irradiation, 16 times;  after the third irradiation, 6 times.
    
    Some minor bands show fluctuations in their appearance before and after irradiation. The 950 cm$^{-1}$  band occurs three times before irradiation, four times after the first irradiation, and nine times after the second irradiation. The 970 cm$^{-1}$ band occurs four times before irradiation and five times after the second and third irradiations. The 910 cm$^{-1}$ band occurs two times before irradiation, four times after the first and second irradiations, and two times after the third irradiation.

\subsubsection{Feldspar}

\begin{figure}[h]
        \centering
        \includegraphics[width=\hsize]{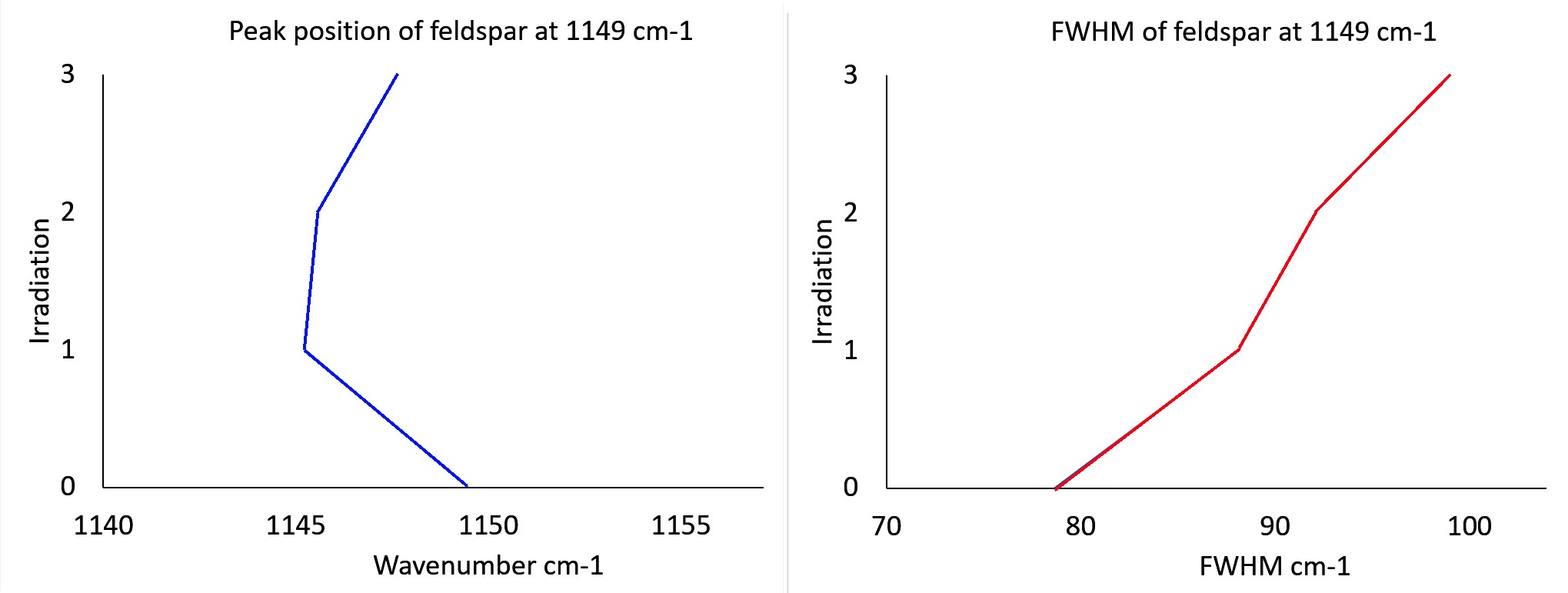}
        \caption{Feldspar 1149 cm$^{-1}$ peak position changes (blue line, left) and FWHM changes (red line, right), with the number of irradiations on the Y-axis.}
        \label{fig:feldspar1149}
    \end{figure}

    In feldspar more bands appear after the third irradiation than before the irradiations, due to polymerization of SiO$_2$ tetrahedra (R1/1-2, 4-7; c1/7-8, b1/-3-5, d1/6-7, a2/2). At some measuring points the bands disappear after the first irradiation and appear after the second irradiation (R1/3-6, b1/4-5, d1/4-6), or only after the third irradiation (c1/7-8, a4b/1-3) (Table SOM II/3). In some cases (at some measurement locations), the bands of feldspar appear after the second and the third irradiations (a1/4-8; see example in Fig. \ref{fig:feldspar1149}). The peak shifts are described in these areas, where the bands appear  before and after irradiation (Tables SOM II/1-2, Table \ref{tab:3}.)

    Both the  major and minor bands appear  mostly before irradiation and after the third irradiation. The disappearance of the minor bands (1000, 980 cm$^{-1}$) is seen after the first and the second irradiations. However, the minor band at 750 cm$^{-1}$ appears more times after the first and the second irradiation, and disappears after the third irradiation. The band at 630-650 cm$^{-1}$ disappears after the first and second irradiations, but appears after the third irradiation. The 950 and 1000 cm$^{-1}$ bands appear  again after the third irradiation (Tables SOM II/1-3,Table \ref{tab:3}) 

\subsubsection{Pyroxene}

    Pyroxenes contain  generally more bands before irradiation, of which 30\% disappear by the end of the irradiation series (two bands: R1/1-2, d1/1-3; all bands: (R1/3-7, c1/1, a4b/1-3). Most bands of pyroxene appear  more times after than before the second irradiation (see Table SOM III/3 (c1/1, 4, 5, 6, b1/2 , d1/7, a2/1), and in some cases more appear  after than before the third irradiation (see Table SOM III/3  b1/4-5), indicating the  formation of a new mineral phase (probably after olivine) \citep{Delvigne1979}. After the final irradiation, in most cases fewer pyroxene bands emerge  than before irradiation, indicating the degradation of the crystal lattice after irradiation  (Table SOM III/3 , last column). This feature indicates a distortion of the crystal lattice, depolimerization of SiO$_4$ tetrahedra \citep{Sharp2006}. Only the band at 700 cm$^{-1}$ of all pyroxene bands shows a continuously disappearing trend: before irradiation it appears 13 times (30\%), but after the third irradiation this band appears only 3 times (0.6\%). However, in several cases the pyroxene appears more times after the third irradiation (Table SOM III/3, a1); this indicates that  a new mineral phase is formed, but   is less crystallized than before the irradiation, indicated by the minor bands appearing fewer times after the irradiation (Tables SOM III/1-3, Table \ref{tab:4}  last column, e.g., c1/2-3, d1/1-4, a2/1-2 Fig. \ref{fig:DRIFT} (a2/1)).

    \begin{figure}[h]
        \centering
        \includegraphics[width=\hsize]{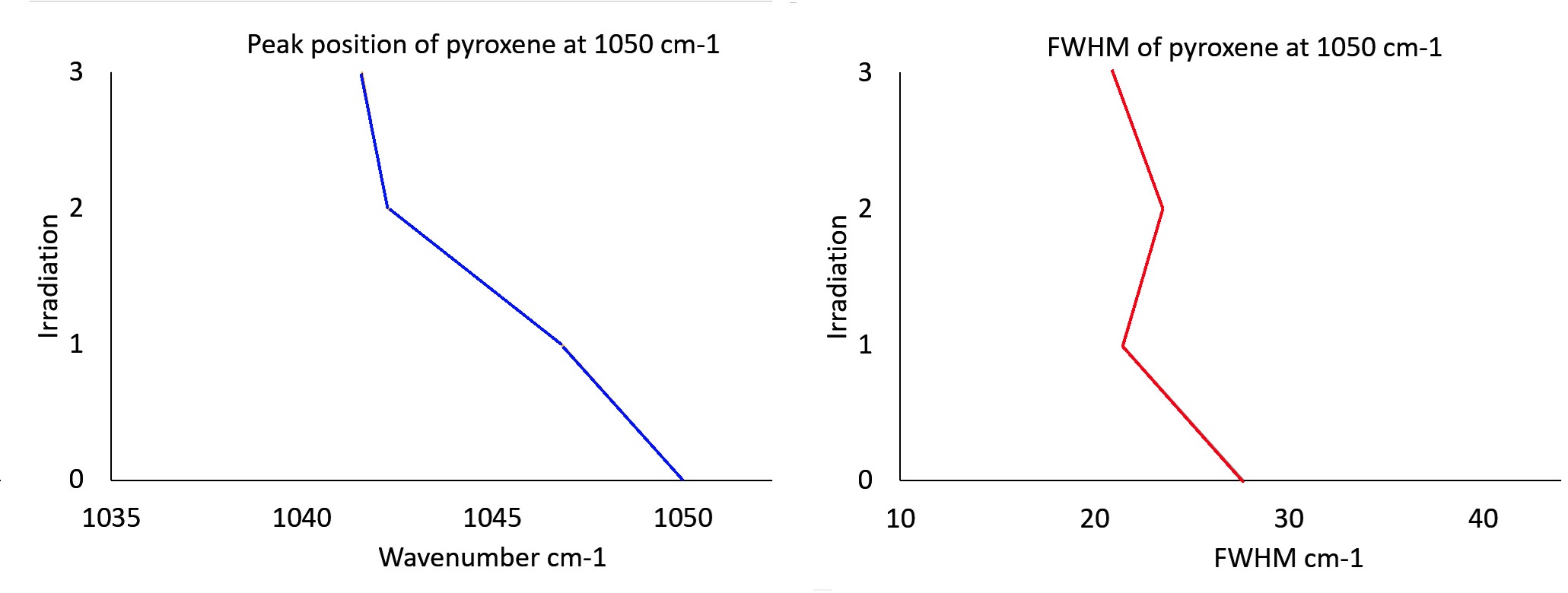}
        \caption{Pyroxene 1050 cm$^{-1}$ peak position changes (blue line, left) and FWHM changes (red line, right), with the number of irradiations on the Y-axis.}
        \label{fig:pyroxene1050}
    \end{figure}
   
    At some measurement locations all bands disappear after the first irradiation: (d1/6-10, a4b/1-3, R1/8-10). In other cases, only one band appears before irradiation, and then disappears after the first irradiation. However, more bands appear after the second irradiation (R1/4-6, b1/2, d1/7), due to the dimerization of SiO$_4$ tetrahedra. The bands of pyroxene (1050, 910, 670 cm$^{-1}$) appear fewer times after the first irradiation, but the disappearing bands partly reappear after the second irradiation (see     1050 cm$^{-1}$   in Fig. \ref{fig:pyroxene1050}). 

\subsubsection{Spinel}

    The band of spinel appears in the range 664-680 cm$^{-1}$. In some cases, spinel is detected only after irradiation (Table SOM IV/3, R1/1-8, a2-1,), but in other cases it does not appear after the third irradiation (c1/1-3). Spinel occurs both before and after irradiation in the d1 chondrule (d1/1-5), and in the a5/1 area (Table \ref{tab:5}).

    \begin{figure}[h]
        \centering
        \includegraphics[width=\hsize]{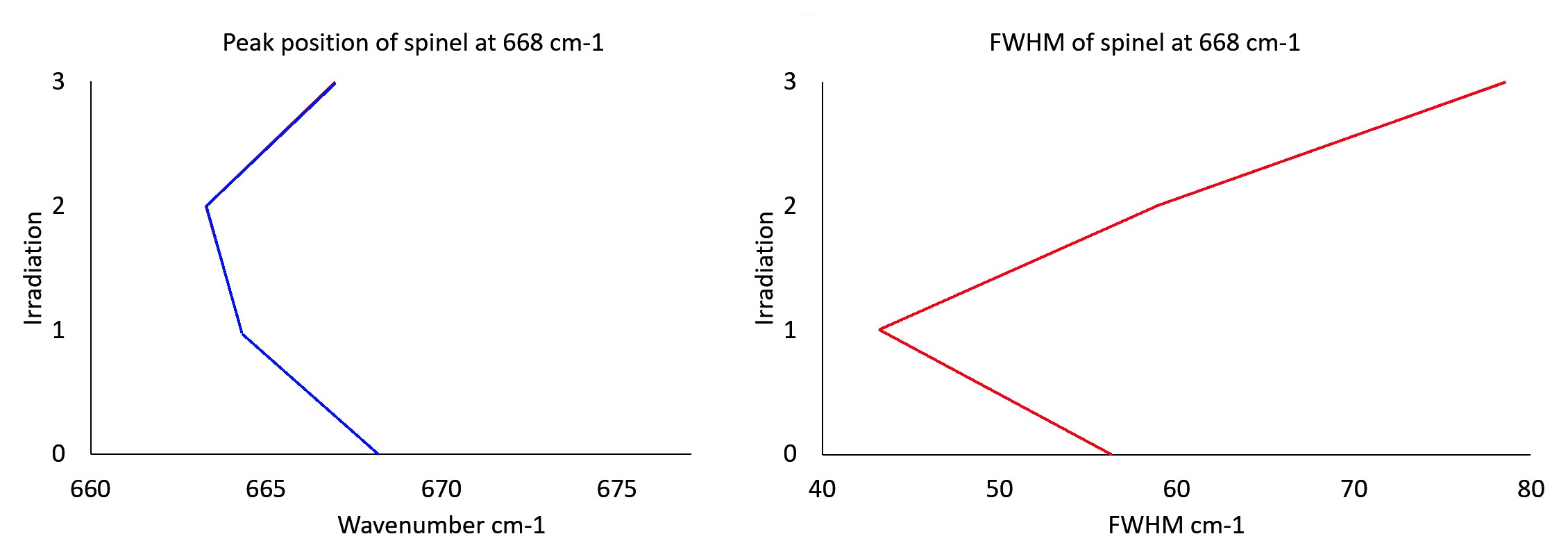}
        \caption{Spinel 668 cm$^{-1}$ peak position changes (blue line, left) and FWHM changes (red line, right), with the number of irradiations on the Y-axis.}
        \label{fig:spinel1668}
    \end{figure}
    
    The major band of spinel at 670 cm$^{-1}$ appears 38 times before irradiation, 18 times after the first irradiation, 35 times after the second irradiation, and 41 times after the third irradiation (Table SOM IV/3). The peak shift of spinel  shows mostly negative values after the first irradiation, but the peak shift changes to a positive trend after the second and third irradiations (Fig. \ref{fig:spinel1668}, Tables SOM IV/1-2 Table \ref{tab:5} ). Spinel shows a decreased   FWHM after the first irradiation (in more measuring points, see Table \ref{tab:6} and the SOM tables), which changes to a positive trend after the second and third irradiations. 

\subsection{Spectral curve pair examples}

    \begin{figure}
   \centering
   \includegraphics[width=\hsize]{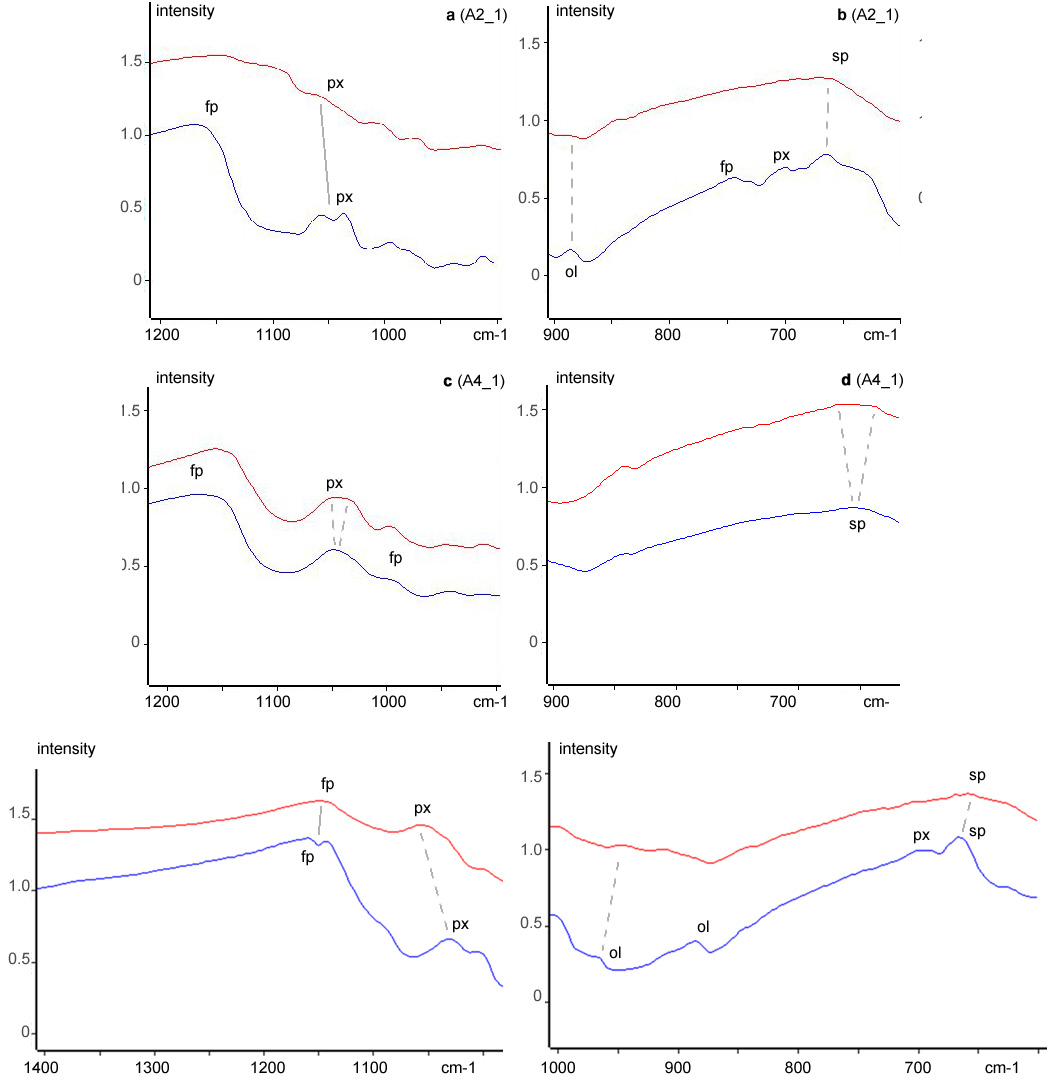}
      \caption{Spectral curve pair examples   demonstrating the changes before (blue, bottom) and after (red, top) the strongest irradiation. The main mineral peaks are labeled with letter pairs.}
         \label{fig:spectralPairs}
   \end{figure}

   Several  examples based on the spectral curves  of the observed irradiation-induced changes are visible in Figure \ref{fig:spectralPairs}, where the blue (lower) curve is the original (nonirradiated) state, while the upper (red) curve is the most heavily irradiated state. The differences between the original and irradiated states can be seen with the following features:

   \begin{itemize}
       \item disappearance of feldspar, and substantial weakening of pyroxene band (A21 location);
       \item disappearance of feldspar and pyroxene, weakening of olivine and spinel bands (A21 location);
       \item peak shift and increased   FWHM of pyroxene (A41 location);
       \item increased FWHM  of spinel (A41 location);
       \item merging of two major bands of feldspar into a single band, and a shift of the pyroxene major band to a higher wavenumber with increased   FWHM after   irradiation (A52 location);
       \item disappearance of olivine main bands, disappearence of  the 700 cm-1 minor band of pyroxene, weakening of spinel after irradiation (A52 location).
   \end{itemize}

\subsection{FWHM observations}

    Only  feldspar shows an increased  FWHM  of the major bands after all the irradiations (Table \ref{tab:6}), while the other minerals show different and variable trends. Pyroxene shows decreased FWHM values after the third irradiation in the case of the 1050 cm$^{-1}$ band (SOM tables), whereas after the first and the second irradiations it shows rather increased FWHM values. Olivine, at the 887 cm$^{-1}$ band, shows a decreased  FWHM after the first and the second irradiations, changing to a moderate increase after the third irradiation. However, at the other major band of olivine: 849 cm$^{-1}$ the FWHM shows an increase after the first irradiation, changing to a decrease after the second and the third irradiations. Consequently, a clear signature of disordering of the crystalline lattice can be observed only in feldspar and in the case of the other minerals the FWHM related changes are not enough for a firm conclusion. 

    \begin{table*}[h]
    \caption{Summary of FWHM main bands}             
    \label{tab:6}      
    \centering                          
    \begin{tabular}{c c c c c c c c}        
    \hline\hline                 
    mineral & spectrum & band & before irr. & 1st irr. & 2nd irr. & 3rd irr. & avg. after irr. \\    
    \hline                        
    spinel & average & 668 & 56.3 & 43.2 & 57.7 & 79.2 & 60 \\ 
spinel & min & 660 & 7 & 7 & 5 & 17 & 9.7 \\ 
spinel & max  & 679 & 130 & 140 & 150 & 130 & 140 \\ 
spinel & median & 667 & 70 & 15.5 & 50 & 80 & 48.5 \\ 
 & average & 668.5 & 65.8 & 51.4 & 65.7 & 76.6 & 64.6 \\ 
pyroxene & average & 1047.5 & 27.7 & 21.4 & 21.5 & 20.3 & 21.1 \\ 
pyroxene & min & 1035 & 5 & 7 & 4 & 5 & 5.3 \\ 
pyroxene & max & 1065 & 54 & 40 & 40 & 40 & 40 \\ 
pyroxene & median & 1046 & 30 & 20 & 23.5 & 20 & 21.2 \\ 
 & average & 1048.4 & 29.2 & 22.1 & 22.3 & 21.3 & 21.9 \\ 
feldspar & average & 1152.12 & 78.7 & 88.2 & 92.1 & 99.2 & 93.2 \\ 
feldspar & min  & 1116 & 12 & 42 & 7 & 70 & 39.7 \\ 
feldspar & max & 1183 & 130 & 130 & 150 & 130 & 136.7 \\ 
feldspar & median & 1148 & 85 & 85 & 90 & 100 & 91.7 \\ 
 & average & 1149.8 & 76.4 & 86.3 & 84.8 & 99.8 & 90.3 \\ 
olivine & average & 887.75 & 11.9 & 8.4 & 6.3 & 12 & 8.9 \\ 
olivine & min & 883 & 7 & 5 & 5 & 3 & 4.3 \\ 
olivine & max & 894 & 18 & 15 & 8 & 30 & 17.7 \\ 
olivine & median & 885.5 & 11 & 5 & 6.5 & 3 & 4.8 \\ 
 & average & 887.6 & 12 & 8.4 & 6.5 & 12 & 8.9 \\ 
olivine & average & 849.2 & 9.7 & 17.5 & 6.1 & 4.3 & 9.3 \\ 
olivine & min & 849 & 5 & 16 & 4 & 3 & 7.7 \\ 
olivine & max & 850 & 16 & 19 & 12 & 5 & 12 \\ 
olivine & median & 849 & 8.5 & 17.5 & 5 & 5 & 9.2 \\ 
 & average & 849.3 & 9.8 & 17.5 & 6.8 & 4.3 & 9.5 \\ 
    \hline                                   
    \end{tabular}
    \end{table*}

    Spinel shows a decreased FWHM after the first irradiation, changing to a positive trend after the second and the third irradiations. The Fe-bearing minerals (spinel, olivine, pyroxene) show a decreasing trend on average after the irradiations compared to pre-irradiation states. Only feldspar shows an increased FWHM of its major band by calculation of the average FWHM of all of irradiations (Table \ref{tab:6}). Pyroxene shows a continuous negative trend in FWHM values. Feldspar shows a positive trend in FWHM due to the deformation of crystal structure after irradiations. The 887 cm$^{-1}$ band of olivine shows a negative trend after the first and second irradiations, changing to a moderate increase after the third irradiation. However, for the other major band at 849 cm$^{-1}$ the FWHM shows an increase after the first irradiation, which changes to a decrease after the second and third irradiations.

    The major bands of olivine (844 and 888 cm$^{-1}$) and feldspar appear  before and after the first and second irradiations. The major band of pyroxene (1050) appears before and after all of the irradiations, except spectrum 2, which does not appear after the second irradiation (Table SOM 0/1, Tables SOM V/1-2).
    
\subsection{General trends}

    Below we evaluate the peak position and FWHM changes together  of several minerals. The Fe-bearing minerals (spinel, olivine, pyroxene) show an average decreasing trend in band positions after irradiation. Not all mineral bands appear after all of the irradiation actions, and so the calculated average peak positions and FWHM values are indicated only in those cases where the given peaks emerged after all the irradiations. 

    Olivine: The major 844 cm$^{-1}$ band shows a negative trend in the peak shift, but an increased  FWHM, while the other major band at 888 cm$^{-1}$ appears only after the second irradiation for most DRIFTS data. The olivine minor band at 987 cm$^{-1}$ appears only before irradiation.  The other minor bands of olivine below 600 cm$^{-1}$ (which could be detected only by DRIFTS) occur after all of the irradiations. The 474 cm$^{-1}$ band shows a decreasing peak shift and an increasing FWHM after the first irradiation, but after the second irradiation this trend inverts (showing negative correlation, i.e., increasing peak shift and decreasing FWHM). The 499 cm$^{-1}$ band shows a positive trend in peak shift and FWHM, but after the second irradiation a negative trend can be observed in both values (Tables SOM V/1-2, Table SOM 0/1). 
    
    The pyroxene major band around 1070 cm$^{-1}$ appears after all of the irradiation actions, and the peak shift continuously decreases after each irradiation. The change in FWHM shows a positive trend after the first and second irradiations (increasing FWHM values), changing to a negative trend after the third irradiation (decreasing FWHM values). Near   this position the band of feldspar at 1161 cm$^{-1}$ splits into two bands (1155 and 1188 cm$^{-1}$) after the third irradiation, which may influence the major band of pyroxene at 1070 cm$^{-1}$ because of the sloped baseline. The minor band at 952 cm$^{-1}$ appears only after the irradiations. This band shows a lower peak position (wavenumber) and FWHM shift after the second irradiation than after the first irradiation. It is characterized by a lower peak shift and higher FWHM value after the third irradiation  compared with the second irradiation.  
    
    The major band of feldspar shows a split of peaks into two bands: 1151 and 1184 cm$^{-1}$ before and after the third irradiation. After the first and the second irradiations the band appears near 1149 cm$^{-1}$. The band at 1150 cm$^{-1}$ shows a negative trend, whereas the 1184 cm$^{-1}$ band shows a positive trend in peak and FWHM shift after the third irradiation. The minor bands at 1000 and 640 cm$^{-1}$ appear only after the first and the third irradiations. Near the 1000 cm$^{-1}$ band, the  950 cm$^{-1}$ band (pyroxene) and the  987 cm$^{-1}$ band (olivine) occur, which may disturb and overlap this band. The 640 cm$^{-1}$ band may be disturbed by the Fe-Mg band of pyroxene at 665 cm$^{-1}$ causing the  disappearance of the 640 cm$^{-1}$ band. The peak position and FWHM of the 640 cm$^{-1}$ band after the third irradiation increases, whereas in the case of 1000 cm$^{-1}$ band only the FWHM shows an increase after the third irradiation.
    
    Using DRIFTS observations, the highest number of spectral changes can be observed after the first irradiation (decrease in  intensity, and shift of bands; see Table SOM 0/1 yellow rows, Tables SOM V/1-2). The   pyroxene band at 1050 m$^{-1}$,  the feldspar band at  1150 cm$^{-1}$, and the olivine band at  844 cm$^{-1}$  decrease in wavenumber. The major bands of olivine show increasing FWHM values, whereas the FWHM of feldspar and pyroxene show variable trends. After the first irradiation, alteration to a metastable phase may have happened, a partly similar situation to that after a shock effect when the loading time is too short (a few seconds), which does not produce  a final structural disordering, but only a metastable structure \citep{Sharp2006}.

\section{Discussion} 

    In this section first we outline how mineral identification was possible during the irradiation series according to the major bands (see Table \ref{tab:8}); the minor bands were more changeable and appeared or disappeared in a somewhat stochastic manner. Overlapping of the bands also caused some difficulties: the major band of spinel overlaps the 670 cm$^{-1}$ band of pyroxene, and the minor 970-980 cm$^{-1}$ band of olivine coincides with pyroxene and feldspar. The problem of overlapping bands was described by \citet{Brunetto2009} in the range 630-680 cm$^{-1}$. Where mineral bands are very close to each other in wavenumber, they may merge and modify the FWHM values, which  may  increase the difficulty of peak position determination. 

\subsection{Overview of results}

    Table \ref{tab:8} below shows the appearance of mineral bands for all measurements, where the ratio of the mineral appearance for each measurement series is indicated as a  percentage. The major band of feldspar always appears before and after the irradiations. The band of pyroxene appears on average 65\% before and after the first and the third irradiations, while this value for the second irradiation is 76\%. The appearance of the 887 cm$^{-1}$ band in olivine weakens continuously after all the irradiations (38 $\rightarrow$ 27 $\rightarrow$ 12\%). The 849 cm$^{-1}$ band of olivine shows a negative trend (decrease of peak position) after the first and the third irradiations, but this occurs more frequently after the second irradiation. The band of spinel (666 cm$^{-1}$) overlaps the Fe-Mg band of olivine, and hence it is difficult to observe and shows a variable trend. It mostly appears after the second irradiation, and less frequently after the first irradiation. Because of the changes and variability in the observability of certain peaks, in the rest of this work we only consider those cases where all of the analyzed bands remained observable during the whole irradiation and measurement series at the given targeted location, and thus where no close position to other bands, merging of bands, or other unfavorable situations influenced the results.

    For the identification of minerals, the major bands are applicable. The major band of olivine at 849 cm$^{-1}$ almost disappeared after the first irradiation, but after the second irradiation became more frequent. In general, the appearance of bands from feldspar, pyroxene, and spinel changed during the irradiations. According to the last column of Table \ref{tab:8}, smaller peak changes occurred in the case of feldspar, and the major band at 1150 cm$^{-1}$ occurs in all of the measurements. The strongest change in spectra can be observed in the case of the 887 cm$^{-1}$ band of olivine, which shows a negative trend (shift of bands to lower wavenumber) until after the third irradiation. The change in the cases of pyroxene and spinel is smaller, but these occur more frequently in the measurement series after the second irradiation than earlier. 

    In A2-1 (Fig. \ref{fig:DRIFT}), after the first irradiation the spectrum is similar to the spectrum before the irradiation, but after the second and third irradiations the band of feldspar near 1150 cm$^{-1}$ disappears, and the intensity of bands between 600 and 750 cm$^{-1}$ decreases after the second irradiation, and finally disappears after the third. In spectra D1-2 (Table SOMII/3)), new bands appear at around 1000 cm$^{-1}$ after the second and third irradiations due to depolimerization of  SiO$_4$ tetrahedra \citep{Johnson2003, Johnson2007}. \citet{Johnson2003} documented changes in the appearance and position of spectral features with increasing shock impact pressure due to depolymerization (disordering of SiO$_4$ tetrahedra from the crystal lattice forming SiO$_2$ molecules) of the silica tetrahedra, which agrees with our observation as well. In spectra A4-1 (Table SOMIII/3) the 1050 cm$^{-1}$ band shifts to a lower wavenumber after the first and second irradiations (metastable disordering; \citealt{Johnson2003,Johnson2007}), but after the third irradiation this band appears at peak positions similar to before the irradiation series (reorganizing structure, observed by  \citealt{Johnson2003,Johnson2007}).
   Feldspar and pyroxene occur in most spectra. Spinel appears mostly later during the irradiation series, or shows a cyclic appearance (Table \ref{tab:9}).

\begin{itemize}
    \item \textbf{Olivine}: The 849 cm$^{-1}$ band shows a peak position decrease after the first irradiation (negative shift), whereas after the second and third irradiations positive peak shifts were observed. The other major 887 cm$^{-1}$ band shows a higher peak shift (positive trend) after the first and third irradiations, but a decrease after the second irradiation. Thus, a variable peak shift change occurs for the 849 cm$^{-1}$ band, whereas the 884 cm$^{-1}$ band shows a negative (decreasing peak shift) trend after the second and third irradiations. The 849 cm$^{-1}$ and 884 cm$^{-1}$ bands unify to one band near 890 cm$^{-1}$ (d1/1-4 after the first irradiation, b1/5, d1/1-2 third irradiation) due to the distortion of crystal lattice after the irradiations. The increase in peak shift of the main doublet of olivine after the third irradiation indicates Fe loss from the crystalline structure \citep{Lantz2017}. This increase in wavenumber is variable in our measurements. The 890 cm$^{-1}$ band disappears in some cases, but the band at 850 cm$^{-1}$ shows both positive and negative trends in peak shifts. The positive trend of peak shift mainly seen at the end of the irradiation series of this work could arise from the deformation of SiO$_4$ tetrahedra of olivine, in a similar fashion to the observed   peak shift in the G band (C=C of graphite) to higher wavenumber, which was described by \citet{Brunetto2009}. The negative peak shift occurs because of Mg loss from Fe-Mg bearing minerals at the irradiated surface \citep{Brunetto2014, Brunetto2020, Lantz2015, Lantz2017}. 
    \item \textbf{Feldspar} shows a decrease in peak position after the first and second irradiations (on average by -2 cm$^{-1}$, but the lowest values are between -37 and -46 cm$^{-1}$), and this peak shift is accompanied by an increase in FWHM (Table \ref{tab:11}) after the second irradiation, averaging 86 cm$^{-1}$), due to crystal lattice defects.
    \item The main band of \textbf{pyroxene} (1047 cm$^{-1}$) shows an average decrease in peak shift (negative trend) after the first irradiation, which becomes an increase (positive trend) after the second and third irradiations.
    \item \textbf{Spinel} shows a decrease in peak shift (negative trend) after the first irradiation, and an increase in peak shift (positive trend) after the third irradiation.
\end{itemize}
  
    After the first irradiation, a negative peak shift was present in Fe-bearing minerals (olivine, pyroxene, spinel) based on peak changes of bands between 630-680 cm$^{-1}$. The positive peak shift of Fe-Mg-bearing minerals after the third irradiation indicates Fe loss. According to \citet{Klima2007} the IR bands of M1-M2 (Fe$^{2+}$) positions are located at 1.2 $\mu$m, 1 $\mu$m, and 2 $\mu$m for silicates (especially pyroxene), but these wavenumbers are too high to be analyzed by the instrument used in this work. \citet{Hanna2020} indicate that the Mg bands of silicates are in the range 715-550 cm$^{-1}$. In our case, this band is at 660-680 cm$^{-1}$, which is the major band of spinels, but it occurs in olivine and pyroxenes as well. Hence, Table \ref{tab:12} shows not only the peak shift of spinels, but also the peak shift of the Fe-Mg band of pyroxene and olivine in general.

    Peak shifts of olivine for the 887 cm$^{-1}$ band are negative on average after the first and second irradiations (-4.3 -- -4.8 cm$^{-1}$ with variation between -15 and +6 cm$^{-1}$), turning to positive after the third irradiation (on average 4.3 cm$^{-1}$)(Table \ref{tab:12}). Hence, the negative peak shifts indicate Mg-loss from the crystal structure after the first and second irradiations. The negative trend of the peak shift after irradiation is mentioned by \citet{Lantz2013, Lantz2015, Lantz2017, Lantz2014}after the first and second irradiation.  

    Cation loss (olivine, feldspar, pyroxene, spinel after the first irradiation; see Table \ref{tab:10}, column 1, negative peak shifts) also causes the disappearance of the major doublet of olivine by the merging of the doublet band to a single band. This one direction change in IR bands was observed in several cases in our measurements after the irradiations (see Table SOM I/5 d1/2 4; second irradiation, b1/5 third irradiation, d1/1-2 third irradiation, a1/6 after the third irradiation). The reverse transformation (splitting to a doublet) could be not observed.

    \begin{table*}[h]
    \caption{Summary peak shift table of the minerals}             
    \label{tab:12}      
    \centering                          
   \begin{tabular}{c c c c c c c}
    \hline\hline
   mineral & spectrum & band & 1st irr. & 2nd irr. & 3rd irr. & avg. peak shift cm$^{-1}$ \\ 
   \hline
    spinel & average & 668 & -3.5 & -3.21 & 5.15 & -0.52 \\ 
    spinel & min & 660 & -15 & -19 & -3 & -12.333 \\ 
    spinel & max  & 679 & 3 & 3 & 11 & 5.666 \\ 
    spinel & median & 667 & -2 & 0 & 4 & 0.666 \\ 
     & average & 668.5 & -4.3 & -4.803 & 4.29 & -1.6300 \\ 
    pyroxene & average & 1047.5 & -2.5 & 0.333 & 2.13 & -0.0296 \\ 
    pyroxene & min & 1035 & -30 & -8 & -23 & -20.333 \\ 
    pyroxene & max & 1065 & 16 & 38 & 30 & 28 \\ 
    pyroxene & median & 1046 & -1 & -6 & 0 & -2.333 \\ 
     & average & 1048.3 & -4.38 & 6.08 & 2.28 & 1.326 \\ 
    feldspar & average & 1152.1 & -2 & -2 & -3.92 & -2.64 \\ 
    feldspar & min  & 1116 & -46 & -37 & -42 & -41.666 \\ 
    feldspar & max & 1183 & 29 & 33 & 26 & 29.333 \\ 
    feldspar & median & 1148 & -7 & 3 & 0 & -1.333 \\ 
     & average & 1149.8 & -6.5 & -0.75 & -4.98 & -4.076 \\ 
    olivine & average & 887.7 & 4.4 & -4 & 0.6 & 0.333 \\ 
    olivine & min & 883 & -2 & -7 & -4 & -4.333 \\ 
    olivine & max & 894 & 6 & 3 & 7 & 5.333 \\ 
    olivine & median & 885.5 & 6 & -4 & 0 & 0.666 \\ 
     & average & 887.56 & 3.6 & -3 & 0.9 & 0.5 \\ 
    olivine & average & 849.2 & -11.3 & 2 & 1.8 & -2.511 \\ 
    olivine & min & 849 & -15 & -15 & -4 & -11.333 \\ 
    olivine & max & 850 & -4 & 15 & 11 & 7.333 \\ 
    olivine & median & 849 & -15 & 4 & 0 & -3.666 \\ 
     & average & 849.3 & -11.33 & 1.5 & 2.2 & -2.544 \\   
    \hline 
    \end{tabular}
    \tablefoot{Min is the lowest value; max is the  highest value.}
    \end{table*}

    The peak position of spinel decreases after the first and second irradiations and increases after the third irradiation (on average -3.5, -3.1 cm$^{-1}$ after the first and second irradiation, 5 cm$^{-1}$ after the third irradiation; see Table \ref{tab:12}). Similar trends can be observed in the case of pyroxenes: -2.5, -0.5 cm$^{-1}$ peak position change after the first and second irradiations, 2 cm$^{-1}$ after the third irradiation. On average a similar trend was observed in feldspar, pyroxene, and spinel, but with a certain deviation (Table SOM II/1, Table SOM III/1). The peak shifts of feldspar show a continuous trend after all of the irradiations, presenting a decreasing wavenumber (negative) trend. The 887 cm$^{-1}$ band of olivine shows a higher peak shift after the first irradiation, that changes to a negative trend after the second irradiation. The 849 cm$^{-1}$ band shows a -11 cm$^{-1}$ peak shift after the first irradiation, which changes to a positive trend after the second and third irradiations.

    Consequently, negative peak shifts occurred after the first irradiation in Fe-Mg bearing minerals. After the second irradiation this trend was variable, but finally after the third irradiation the average values of peak position shift of olivine turned to positive. All of these happened in the top less than micrometer thick layer of the sample, where the maximum damage occurs at a shallow penetration depth \citep{Demyk2004}. In the case of feldspar, on average the negative peak shift changed to positive after the second irradiation, but after the third irradiation these values turned negative again. Pyroxene shows negative values in peak shift after the first and second irradiations, which change to positive after the third irradiation. 

\subsection{Comparison of the results with other works}

    In this section we present the discussion of the mineral changes   in the following order: 1) summary of the related results of other published research projects (there might be a small overlap with those that are listed in the Introduction; however, they are presented here as well for     comparison); 2) comparison of our results with those of others regarding the peak changes (position shift and FWHM); 3) presentation of the possible reasons for these changes.
    
    Published works by other authors present observations in the far-IR and mid-IR regions after the irradiation of minerals, focusing on measured peak shifts, but not FWHM values. In general, other works show a decrease in intensity and increase in spectral slope (known as darkening) in the IR spectrum \citep{Strazzulla2005}. Peak shifts of olivine and pyroxene are characteristically in the range 0.4-0.12 $\mu$m (1-25
    cm$^{-1}$) \citep{Lantz2015, Lantz2017} after a flux of $\sim$10$^{13}$ ions (sec$^{-1}$ cm$^{-2}$) up to a total fluence of $\sim$6 $\times$10$^{16}$ ions cm$^{-2}$, indicating a variation in peak shift with increasing irradiation energy. Peak shifts are negative when they are defined in wavenumbers, and each band is influenced differently, but all of the spectra are affected \citep{Brunetto2014}. They observed that the shift of the $\sim$870 cm$^{-1}$ band is much larger (12 cm$^{-1}$) than the shifts possibly caused by the heterogeneity of the matrix olivines. Incorporation of nuclei due to bombarding with bigger ions (H$^+$, He$^+$, Ar$^+$, Ar$^{2+}$) at higher energies (60-400 kev) distorts the crystal lattice more. In our case, the irradiation energy is 1 keV with increasing fluences after each irradiation, and the ion size is smaller as we used H$^+$ ion with fluences of 10$^{11}$, 10$^{14}$, and 10$^{17}$ ions/sec/cm$^{-2}$. Instead, the flux of the second irradiation was similar to several tests described in the literature. The peak shift of the olivine band (887 cm$^{-1}$)  was a negative value (-4 cm$^{-1}$), but with high variability (-15 and +15 cm$^{-1}$). According to \citet{Brunetto2006}, the silicates (pyroxenes and olivine) show progressive amorphization due to elastic collision of ions,
causing a shift  of bands to lower wavenumbers and decrease in band intensity (amorphization depends on ion size:  amorphization is higher by bombardment of Ar$^+$ than H$^+$). Similar band changes were observed in our olivine, pyroxene, and  spinel spectra, but with a higher variation: between -19 and 11 for spinel; between -23 and 30 for pyroxene; between -7 and +7 (887cm$^{-1}$) and between  -15 and +15 (850 cm$^{-1}$) for olivine. 
    
    Our results on peak shift of the 849 cm$^{-1}$ olivine band were negative after the first irradiation, which turned to a positive trend after the  third and second irradiations (Table \ref{tab:10}). This observed negative peak shift is similar to the results of \citet{Lantz2015, Lantz2017} due to Mg-loss from the crystal lattice on the irradiated surface. The other major band of olivine (887 cm$^{-1}$) shows a partly opposite trend: on average a positive value (increase in wavenumber) after the first irradiation, which turns   negative after the second irradiation, but after the third irradiation on average remains almost zero. The negative peak shift after the first and second irradiations is similar to that found by  \citet{Lantz2015, Lantz2017}. The peak shift of feldspar is negative after the first and third irradiations, and turns positive after the second irradiation, which is difficult to explain. Pyroxene and spinel have negative peak shift values after the first irradiation, turning positive after the second and third irradiations. The negative peak shift of pyroxene is similar to other data \citep{Lantz2015, Lantz2017}, but the positive peak shift has not been described by others after irradiation. However, \citet{Johnson2003} observed a peak shift due to distortion of SiO$_4$ tetrahedra of feldspar due to shock deformation of the crystal lattice, and hence in our case the positive peak shift could be interpreted as crystal lattice defects due to irradiation.
    
    Considering the IR spectroscopic data presented in the literature from others \citep{Lantz2015, Lantz2017, Brunetto2018, Brunetto2020}, the spectral changes after irradiation are rarely studied by Raman spectroscopy. However, the disordering of meteorite minerals after shock metamorphism is usually studied by this Raman spectroscopy, which may be used as a reference for this work evaluating crystalline structural changes. \citet{Brunetto2014} observed the merging of olivine Raman doublet bands (820, 850 cm$^{-1}$) to a single band at around 840 cm$^{-1}$ due to irradiation. Similarly to their results, after the irradiation we observed one unified band near 890 cm$^{-1}$. In our data the other major olivine band disappeared at 850 cm$^{-1}$ after the first irradiation (d1/2-4) and the third irradiation (b1/5, d1/1-2), and after the second irradiation (c1/2-3). The olivine disappeared in most cases (a5/2, an4b/1-3, a2/2, a3/2, c1/7-9) after all of the irradiations. \citet{Brunetto2014} explains the disappearance of the olivine major band by the deformation of crystal structure due to the irradiation, when the depletion of Mg ions from the irradiated surface layer of olivine occurred.
    
    An important problem in the interpretation of band changes is the possibility of overlapping IR bands affecting the FWHM values, mostly observed in the region around 650 cm$^{-1}$. This effect is also observed by \citet{Brunetto2014} who suggested surveying with more detectors or at higher spectral resolution in the future. In our case, the Fe-Mg bands of olivine, pyroxene and spinel are near to each other, and in the reflection spectra the FWHM is also influenced by the characteristic of the baseline (Fig. \ref{fig:DRIFT}) making the FWHM interpretation difficult in certain cases.
    
    \citet{Brunetto2020} studied bulk meteorite samples using He ion irradiation (fluences -20-40 keV 6$\times$10$^{16}$ - 10$^{13}$/ions/sec), but they had FTIR bands in the far IR region, below 600  cm$^{-1}$ (a region which cannot be detected by the Vertex Hyperion 2000 IR-reflection microscope used in this work). They measured peak shift values varying between -1.5 and -13.5 cm$^{-1}$. Our bands in DRIFTS spectra appear at 462-478 cm$^{-1}$, and in the range 489-503 cm$^{-1}$, but characterized by variable peak shift values (-45-37 cm$^{-1}$ after the first irradiation and between -34 and -7 cm$^{-1}$ after the second irradiation). Only after the third irradiation do the emerged peak shifts differ slightly from those of \citet{Brunetto2020} (1 and -7 cm$^{-1}$). The negative peak shifts are similar to those found by  \citet{Brunetto2020}, but the positive peak shifts we observed are not described in the literature. However, their measured negative shift is characterized by a smaller range (-1.5 and -13.5) than our data. According to \citet{Lantz2015, Lantz2017}, gradually increasing negative peak shift values (decreasing trend) occur in the MIR region due to Mg loss from the structure, with band position versus band shift as follows: 980 cm$^{-1}$ with -54 cm$^{-1}$, 960 cm$^{-1}$ with -92 cm$^{-1}$, 950 cm$^{-1}$ with -60 cm$^{-1}$, 1020 cm$^{-1}$ with -40 cm$^{-1}$, and 867 cm$^{-1}$ with -2 cm$^{-1}$. In our case the peak shift values are variable, showing both negative and positive values (between -12 and +6 cm$^{-1}$ for olivine, -46 and +33 cm$^{-1}$ for feldspar, and -30 and 38 cm$^{-1}$ for pyroxene). Average peak shift values are negative after the first and second irradiations, but turn positive after the third irradiation. In summary, we observed trends in peak shifts similar to  \citet{Lantz2015, Lantz2017} indicating Mg loss from the irradiated surface of the crystal lattice at the first and second (weaker) irradiations;   later at the much harder third irradiation, the positive peak shift arises from the amorphization (i.e., falling apart) of the tetrahedra structures.
    
    Because the published works do not present FWHM values after irradiations, we used shock metamorphism induced mineral deformation for comparison, as the crystalline lattice was deformed there as well. In general, the increase in FWHM in  both  Raman and IR spectroscopy indicates crystal lattice deformation by shock effects \citep{Sharp2006}, and thus deformation of the original well-ordered crystalline structure. Unfortunately, only a small amount of  Raman data are available from other works after irradiation experiments. In relation to Raman spectroscopy, a small amount of data is mentioned by \citet{Lantz2017} and \citet{Brunetto2009} on the FWHM of IR G band (C=C) of graphite showing a poorly ordered pristine structure after the irradiation, indicating disordering of the crystalline structure.

\subsection{Suggestions for future tests}    

    Although several published works present IR peak position change by irradiation  \citep{Lantz2015, Lantz2017, Brunetto2014, Brunetto2020}, unfortunately FWHM calculations are not presented by other authors. We provide several missing data points, mainly related to Fe-Mg bearing minerals (pyroxene, olivine) and feldspar. In future tests the level of compositional change should be followed by more focused analysis.

    Because the variation in the appearance of minor bands is high (variable appearance and disappearance), we present the major bands (845, 887 cm$^{-1}$ for olivine, 1050 cm$^{-1}$ for pyroxene, 1150 cm$^{-1}$ for feldspar), which characterize the changes. In the case of spinel, only the major band between 660-680 cm$^{-1}$ appeared, and hence only this band can be presented. It is reasonable to further analyze the main band positions, partly as they appear better (longer) preserved even during strong irradiation and might be identifiable even at heavily space weathered asteroid surfaces as well.

    Because of the overlapping of mineral bands, further measurements of spinel by other methods (microXRD, Raman) is important. New microXRD measurements could also help to distinguish the appearance of new phases and disappearance of minerals after irradiation. 

\section{Conclusion}

    In this work IR reflection and DRIFTS measurements were carried out on the NWA10580 CO3 meteorite after three proton irradiation actions  to see the alteration of crystal structure (change in peak positions and FWHM values) in order to simulate space weathering effects. This work used proton irradiation with an average 1 keV solar wind ion energy in three  gradually increased doses, using 10$^{11}$ ion/cm$^2$, 10$^{14}$ ion/cm$^2$, and 10$^{17}$ ion/cm$^2$ fluent values.
    
    The changes observed after the irradiations and the listed peak positions are also important themselves as many of them have not been published previously. Minor bands presented variable appearance and disappearance from the irradiations. Thus, we focused on the more obvious major bands for mineral identification in the  cases where the same minerals were identified firmly after all irradiations. These observations mostly presented a negative peak shift after the first and second irradiations in pyroxene and feldspar (-31 and -45 cm$^{-1}$), as in  the literature \citep{Brunetto2014, Lantz2015, Lantz2017}, caused mainly by Mg loss. However, after the third irradiation, positive values in wavenumber change emerged for all silicates, which may arise from the distortion and breaking down of SiO$_4$ tetrahedra, as was described in the case of shock deformation by \citet{Johnson2003}. We also presented the first published tests of meteorite-based feldspar irradiation analysis, with the splitting of the major band into two bands. 
    
    The FWHM values of the major bands show a positive (increasing) trend after irradiations in the case of feldspars using IR reflection. The increase in FWHM of spinel also indicates that the main structural deformation occurred after the third irradiation. However, the FWHM change in pyroxene is negative (decreases) after the first irradiation, but is positive (increases) after the second and third irradiations. This situation indicates that spinel is more sensitive to irradiation as the crystalline lattice deformation starts already at the weakest first irradiation. The change in FWHM of olivine is negative (decreases) after the first irradiation, and changes to positive values after the second and third irradiations, similarly to pyroxene. 
    
    By the comparison of DRIFTS and reflection IR data, we observed the following trends. The peak position of major mineral bands were at similar wavenumbers in the  DRIFTS and reflection methods, but in minor bands differences could be observed. Spinel could be better identified by IR reflection than by DRIFTS. The olivine band at lower wavenumbers was better detected by DRIFTS because of the limited wavelength coverage (up to 400 cm$^{-1}$ vs 600 cm$^{-1}$). 
    
    The overlapping of the bands influences the band identification and also the FWHM measurements. Overlapping mainly occurred in the regions 650-680 cm$^{-1}$ and 1150-1180 cm$^{-1}$ using IR reflection measurements, while in DRIFTS, the minor bands of olivine (987 cm$^{-1}$) and pyroxene (952 cm$^{-1}$) could disturb each other. The overlap of mineral bands in the region 630-680 cm$^{-1}$ influences the FWHM data in reflection IR measurements: the co-existence of bands of more types of minerals (spinel: 660-680 cm$^{-1}$, feldspar: 640 cm$^{-1}$, Fe-Mg vibration of olivine: 660 cm$^{-1}$, and pyroxene: 670 cm$^{-1}$) decreases the FWHM values (Fig. \ref{fig:DRIFT}).
    
    The overlapping of bands near 650 cm$^{-1}$ can influence the band identification (decreasing the FWHM values in the case of the  co-existence of more Fe-Mg minerals). In reflection IR spectra feldspar has a peak position at 631-640 cm$^{-1}$, while the Fe-Mg vibration of olivine is in the region near 670 cm$^{-1}$, while the  Si-O band of spinel varies in the range 660-680 cm$^{-1}$ in the reflection IR spectra. With increasing irradiation fluences, the doublet of olivine (849 cm$^{-1}$ and 880 cm$^{-1}$) merged into a single band at 887 cm$^{-1}$ due to disordering of SiO$_4$ tetrahedra, in agreement with earlier results \citep{Moort2007}.
    
    Comparing our results with those of  other authors, it is worth mentioning that previous tests used mainly higher energy protons around 30 keV, including \citet{Brunetto2018} and \citet{Strazzulla2005}, while our irradiation was at 1 keV. However, the fluences 5$\times$10$^{14}$, 2$\times$10$^{15}$, 1$\times10^{16}$, and  3$\times10^{16}$ ions/cm$^2$ used by \citet{Brunetto2018} were lower than our third irradiation flux (10$^{17}$ ions/cm$^2$). Our first irradiation was lower than in the above-indicated reference, but the flux of the second irradiation was in   range similar to that used by  \citet{Brunetto2018}. Our results (negative peak shift values in wavenumber) after the first and second irradiations were similar to published references (produced by Mg loss), but the average data of the third irradiation (positive values in wavenumber) is a new finding.  
    
    Considering all our measurements,  a negative peak shift (caused by Mg loss) was first observed, followed by amorphization of the crystal lattice. However distortion of SiO$_4$ tetrahedra was not specifically presented in other works, only a description of amorphization in general. Our results indicate that with further analysis it might be possible to separate different levels of destruction of the crystalline structure and link such information to the spectral properties of asteroids. The observed positive peak shift after our third irradiation was also not mentioned in the references, because in previous works the irradiations were done with lower fluences. With further analysis such IR data might be linked to exposure durations of asteroid surfaces, might confirm or fine-tune space weathering durations, and will help to calculate breakup events using high spatial resolution spacecraft IR data in the future. 
    

\begin{acknowledgements}
      This project was supported by the K\_138594 project of NKFIH. Z.J. is grateful for the support of the Hungarian Academy of Sciences through the János Bolyai Research Scholarship. This work was also supported in part by Europlanet 2024 RI which has received funding from the European Union’s Horizon 2020 Research Innovation Programme under Grant Agreement No. 871149.
\end{acknowledgements}

\bibliographystyle{aa}
\bibliography{references}

\begin{thebibliography}{33}
\expandafter\ifx\csname natexlab\endcsname\relax\def\natexlab#1{#1}\fi

\bibitem[{Biri {et~al.}(2012)Biri, Rácz, \& Pálinkás}]{Biri2012}
Biri, S., Rácz, R., \& Pálinkás, J. 2012, Review of Scientific Instruments,
  83, 02A341

\bibitem[{{Brucato} {et~al.}(2004){Brucato}, {Strazzulla}, {Baratta}, \&
  {Colangeli}}]{Brucato2004}
{Brucato}, J.~R., {Strazzulla}, G., {Baratta}, G., \& {Colangeli}, L. 2004,
  \aap, 413, 395

\bibitem[{{Brunetto} {et~al.}(2018){Brunetto}, {Lantz}, {Dionnet}, {Borondics},
  {Al{\'e}on-Toppani}, {Baklouti}, {Barucci}, {Binzel}, {Djouadi}, {Kitazato},
  \& {Pilorget}}]{Brunetto2018}
{Brunetto}, R., {Lantz}, C., {Dionnet}, Z., {et~al.} 2018, \planss, 158, 38

\bibitem[{{Brunetto} {et~al.}(2014){Brunetto}, {Lantz}, {Ledu}, {Baklouti},
  {Barucci}, {Beck}, {Delauche}, {Dionnet}, {Dumas}, {Duprat}, {Engrand},
  {Jamme}, {Oudayer}, {Quirico}, {Sandt}, \& {Dartois}}]{Brunetto2014}
{Brunetto}, R., {Lantz}, C., {Ledu}, D., {et~al.} 2014, \icarus, 237, 278

\bibitem[{{Brunetto} {et~al.}(2020){Brunetto}, {Lantz}, {Nakamura}, {Baklouti},
  {Le Pivert-Jolivet}, {Kobayashi}, \& {Borondics}}]{Brunetto2020}
{Brunetto}, R., {Lantz}, C., {Nakamura}, T., {et~al.} 2020, \icarus, 345,
  113722

\bibitem[{{Brunetto} {et~al.}(2009){Brunetto}, {Pino}, {Dartois}, {Cao},
  {d'Hendecourt}, {Strazzulla}, \& {Br{\'e}chignac}}]{Brunetto2009}
{Brunetto}, R., {Pino}, T., {Dartois}, E., {et~al.} 2009, \icarus, 200, 323

\bibitem[{{Brunetto} {et~al.}(2006){Brunetto}, {Vernazza}, {Marchi}, {Birlan},
  {Fulchignoni}, {Orofino}, \& {Strazzulla}}]{Brunetto2006}
{Brunetto}, R., {Vernazza}, P., {Marchi}, S., {et~al.} 2006, \icarus, 184, 327

\bibitem[{{Delvigne} {et~al.}(1979){Delvigne}, {Bisdom}, {Sleeman}, \&
  {Stoops}}]{Delvigne1979}
{Delvigne}, J., {Bisdom}, E. B.~A., {Sleeman}, J., \& {Stoops}, G. 1979,
  Pedologie, 29, 247

\bibitem[{{Demyk} {et~al.}(2004){Demyk}, {d'Hendecourt}, {Leroux}, {Jones}, \&
  {Borg}}]{Demyk2004}
{Demyk}, K., {d'Hendecourt}, L., {Leroux}, H., {Jones}, A.~P., \& {Borg}, J.
  2004, \aap, 420, 233

\bibitem[{Drochner \& Vogel(2012)}]{Drocher2012}
Drochner, A. \& Vogel, G.~H. 2012, Diffuse Reflectance Infrared Fourier
  Transform Spectroscopy: an In situ Method for the Study of the Nature and
  Dynamics of Surface Intermediates (John Wiley \& Sons, Ltd), 445--475

\bibitem[{{Dukes} {et~al.}(1999){Dukes}, {Baragiola}, \&
  {McFadden}}]{Dukes1999}
{Dukes}, C.~A., {Baragiola}, R.~A., \& {McFadden}, L.~A. 1999, \jgr, 104, 1865

\bibitem[{{Dukes} {et~al.}(2015){Dukes}, {Fulvio}, \& {Baragiola}}]{Dukes2015}
{Dukes}, C.~A., {Fulvio}, D., \& {Baragiola}, R.~A. 2015, in LPI Contributions,
  Vol. 1878, Space Weathering of Airless Bodies: An Integration of Remote
  Sensing Data, Laboratory Experiments and Sample Analysis Workshop, ed. {LPI
  Editorial Board}, 2063

\bibitem[{{Fraser} \& {Griffiths}(1990)}]{Fraser1990}
{Fraser}, D. J.~J. \& {Griffiths}, P.~R. 1990, Applied Spectroscopy, 44, 193

\bibitem[{{Hamilton}(2010)}]{Hamilton2010}
{Hamilton}, V.~E. 2010, Chemie der Erde / Geochemistry, 70, 7

\bibitem[{{Hanna} {et~al.}(2020){Hanna}, {Hamilton}, {Haberle}, {King},
  {Abreu}, \& {Friedrich}}]{Hanna2020}
{Hanna}, R.~D., {Hamilton}, V.~E., {Haberle}, C.~W., {et~al.} 2020, \icarus,
  346, 113760

\bibitem[{{Hapke} {et~al.}(1975){Hapke}, {Cassidy}, \& {Wells}}]{Hapke1975}
{Hapke}, B., {Cassidy}, W., \& {Wells}, E. 1975, Moon, 13, 339

\bibitem[{{Johnson} {et~al.}(2003){Johnson}, {H{\"o}rz}, \&
  {Staid}}]{Johnson2003}
{Johnson}, J.~R., {H{\"o}rz}, F., \& {Staid}, M.~I. 2003, American
  Mineralogist, 88, 1575

\bibitem[{{Johnson} {et~al.}(2007){Johnson}, {Staid}, \& {Kraft}}]{Johnson2007}
{Johnson}, J.~R., {Staid}, M.~I., \& {Kraft}, M.~D. 2007, American
  Mineralogist, 92, 1148

\bibitem[{{Ka{\v{n}}uchov{\'a}} {et~al.}(2017){Ka{\v{n}}uchov{\'a}}, {Boduch},
  {Domaracka}, {Palumbo}, {Rothard}, \& {Strazzulla}}]{Kanuchova2017}
{Ka{\v{n}}uchov{\'a}}, Z., {Boduch}, P., {Domaracka}, A., {et~al.} 2017, \aap,
  604, A68

\bibitem[{{Klima} {et~al.}(2007){Klima}, {Pieters}, \& {Dyar}}]{Klima2007}
{Klima}, R.~L., {Pieters}, C.~M., \& {Dyar}, M.~D. 2007, Meteorit. Planet.
  Sci., 42, 235

\bibitem[{Korte(1988)}]{Korte1988}
Korte, E.~H. 1988, Infrarot-Spektroskopie diffus reflektierender Proben, ed.
  H.~G{\"u}nzler, R.~Borsdorf, W.~Fresenius, W.~Huber, H.~Kelker,
  I.~L{\"u}derwald, G.~T{\"o}lg, \& H.~Wisser (Berlin, Heidelberg: Springer
  Berlin Heidelberg), 91--123

\bibitem[{{Kortum} \& {Delfs}(1964)}]{Kortum1964}
{Kortum}, G. \& {Delfs}, H. 1964, Spectrochim. Acta, 20, 405

\bibitem[{Lafuente {et~al.}(2016)Lafuente, Downs, Yang, \&
  Stone}]{Lafuente2016}
Lafuente, B., Downs, R.~T., Yang, H., \& Stone, N. 2016, 1. The power of
  databases: The RRUFF project, ed. T.~Armbruster \& R.~M. Danisi (Berlin,
  München, Boston: De Gruyter (O)), 1--30

\bibitem[{{Lantz} \& {Brunetto}(2014)}]{Lantz2014}
{Lantz}, C. \& {Brunetto}, R. 2014, in Asteroids, Comets, Meteors 2014, ed.
  K.~{Muinonen}, A.~{Penttil{\"a}}, M.~{Granvik}, A.~{Virkki}, G.~{Fedorets},
  O.~{Wilkman}, \& T.~{Kohout}, 305

\bibitem[{{Lantz} {et~al.}(2015){Lantz}, {Brunetto}, {Barucci}, {Dartois},
  {Duprat}, {Engrand}, {Godard}, {Ledu}, \& {Quirico}}]{Lantz2015}
{Lantz}, C., {Brunetto}, R., {Barucci}, M.~A., {et~al.} 2015, \aap, 577, A41

\bibitem[{{Lantz} {et~al.}(2017){Lantz}, {Brunetto}, {Barucci}, {Fornasier},
  {Baklouti}, {Bour{\c{c}}ois}, \& {Godard}}]{Lantz2017}
{Lantz}, C., {Brunetto}, R., {Barucci}, M.~A., {et~al.} 2017, \icarus, 285, 43

\bibitem[{{Lantz} {et~al.}(2013){Lantz}, {Clark}, {Barucci}, \&
  {Lauretta}}]{Lantz2013}
{Lantz}, C., {Clark}, B.~E., {Barucci}, M.~A., \& {Lauretta}, D.~S. 2013, \aap,
  554, A138

\bibitem[{{Lazzarin} {et~al.}(2006){Lazzarin}, {Marchi}, {Moroz}, {Brunetto},
  {Magrin}, {Paolicchi}, \& {Strazzulla}}]{Lazzarin2006}
{Lazzarin}, M., {Marchi}, S., {Moroz}, L.~V., {et~al.} 2006, \apjl, 647, L179

\bibitem[{Mitchell(1993)}]{Mitchell1993}
Mitchell, M.~B. 1993, Fundamentals and Applications of Diffuse Reflectance
  Infrared Fourier Transform (DRIFT) Spectroscopy, 351--375

\bibitem[{{Sharp} \& {de Carli}(2006)}]{Sharp2006}
{Sharp}, T.~G. \& {de Carli}, P.~S. 2006, in Meteorites and the Early Solar
  System II, ed. D.~S. {Lauretta} \& H.~Y. {McSween}, 653

\bibitem[{{Strazzulla} {et~al.}(2005){Strazzulla}, {Dotto}, {Binzel},
  {Brunetto}, {Barucci}, {Blanco}, \& {Orofino}}]{Strazzulla2005}
{Strazzulla}, G., {Dotto}, E., {Binzel}, R., {et~al.} 2005, \icarus, 174, 31

\bibitem[{{Van de Moort{\`e}le} {et~al.}(2007){Van de Moort{\`e}le}, {Reynard},
  {McMillan}, {Wilson}, {Beck}, {Gillet}, \& {Jahn}}]{Moort2007}
{Van de Moort{\`e}le}, B., {Reynard}, B., {McMillan}, P.~F., {et~al.} 2007,
  Earth and Planetary Science Letters, 261, 469

\bibitem[{{Vernazza} {et~al.}(2013){Vernazza}, {Fulvio}, {Brunetto}, {Emery},
  {Dukes}, {Cipriani}, {Witasse}, {Schaible}, {Zanda}, {Strazzulla}, \&
  {Baragiola}}]{Vernazza2013}
{Vernazza}, P., {Fulvio}, D., {Brunetto}, R., {et~al.} 2013, \icarus, 225, 517

\end{thebibliography}


\begin{appendix}
    \section{Peak shift and summary tables}
    \subsection{Olivine}

     \begin{table}[h]
    \caption{Peak shift of the major 849 cm$^{-1}$ band after the irradiations.}             
    \label{tab:1}      
    \centering                          
    \begin{tabular}{c c c c c c}        
    \hline\hline                 
    area & spectrum & band & 1st irr. & 2nd irr. & 3rd irr. \\    
    \hline                        
       a3 & 1 & 849 & -15 & 15 & -4 \\ 
a4 & 1 & 849 & -15 & 8 & 3 \\ 
a4 & 2 & 849 & -4 & 0 & 0 \\ 
a5 & 1 & 849 &  & -15 & 11 \\ 
a1 & 4 & 850 &  &  & -1 \\ 
 & average & 849.2 & -11.33 & 2 & 1.8 \\ 
 & min & 849 & -15 & -15 & -4 \\ 
 & max & 850 & -4 & 15 & 11 \\ 
 & median & 849 & -15 & 4 & 0 \\ 
    \hline                                   
    \end{tabular}
    \end{table}

    \begin{table}[h]
    \caption{Peak shift of the major 849 cm$^{-1}$ band of olivine with statistic calculations}             
    \label{tab:2}      
    \centering                          
    \begin{tabular}{c c c c c c}        
    \hline\hline                 
    area & spectrum & band & 1st irr. & 2nd irr. & 3rd irr. \\    
    \hline                        
       c1 & 2 & 894 &   & -7 &   \\ 
c1 & 3 & 894 &   & 3 &   \\ 
c1 & 4 & 894 &   & -7 &   \\ 
c1 & 5 & 894 &   & -7 &   \\ 
b1 & 5 & 887 &   & - & 4 \\ 
d1 & 1 & 884 & 6 & -3 & -4 \\ 
d1 & 2 & 884 & 6 & -3 & -4 \\ 
d1 & 3 & 884 & 6 &   &   \\ 
d1 & 4 & 884 & 6 &   &   \\ 
a2 & 1 & 883 & -2 &   &   \\ 
a1 & 3 & 884 &   & -4 & 7 \\ 
a1 & 6 & 887 &   &   & 0 \\ 
  & average & 887.75 & 4.4 & -4 & 0.6 \\ 
  & min & 883 & -2 & -7 & -4 \\ 
  & max & 894 & 6 & 3 & 7 \\ 
  & median & 885.5 & 6 & -4 & 0 \\ 
    \hline                                   
    \end{tabular}
    \end{table}

    \subsection{Feldspar}

     \begin{table}[h]
    \caption{Peak shift values of the major feldspar band after irradiation}             
    \label{tab:3}      
    \centering                          
    \begin{tabular}{c c c c c c}        
    \hline\hline                 
    area & spectrum & band & 1st irr. & 2nd irr. & 3rd irr. \\    
    \hline                        
       c1 & 2 & 894 &   & -7 &   \\ 
c1 & 3 & 894 &   & 3 &   \\ 
c1 & 4 & 894 &   & -7 &   \\ 
c1 & 5 & 894 &   & -7 &   \\ 
b1 & 5 & 887 &   & - & 4 \\ 
d1 & 1 & 884 & 6 & -3 & -4 \\ 
d1 & 2 & 884 & 6 & -3 & -4 \\ 
d1 & 3 & 884 & 6 &   &   \\ 
d1 & 4 & 884 & 6 &   &   \\ 
a2 & 1 & 883 & -2 &   &   \\ 
a1 & 3 & 884 &   & -4 & 7 \\ 
a1 & 6 & 887 &   &   & 0 \\ 
  & average & 887.75 & 4.4 & -4 & 0.6 \\ 
  & min & 883 & -2 & -7 & -4 \\ 
  & max & 894 & 6 & 3 & 7 \\ 
  & median & 885.5 & 6 & -4 & 0 \\ 
    \hline                                   
    \end{tabular}
    \end{table}

    \subsection{Pyroxene}
     \begin{table}[h]
    \caption{Peak shift of the major band of pyroxene after all irradiations}             
    \label{tab:4}      
    \centering                          
    \begin{tabular}{c c c c c c}        
    \hline\hline                 
    area & spectrum & band & 1st irr. & 2nd irr. & 3rd irr. \\    
    \hline                        
    d1 & 1 & 1039 & 6 & -6 & 0 \\ 
    d1 & 2 & 1039 & 6 & -6 & 0 \\ 
    d1 & 3 & 1054 & -9 & -6 & 0 \\ 
    a2 & 1 & 1065 & -30 & 15 & 11 \\ 
    a4 & 2 & 1039 & -12 & 38 & -4 \\ 
      & average & 1047.2 & -7.8 & 7 & 1.4 \\ 
      & min & 1039 & -30 & -6 & -4 \\ 
      & max & 1065 & 6 & 38 & 11 \\ 
      & median & 1039 & -9 & -6 & 0 \\ 
    \hline                                   
    \end{tabular}
    \end{table}

    \subsection{Spinel}

     \begin{table}[h]
    \caption{Peak shift of band of spinel after all irradiations.}             
    \label{tab:5}      
    \centering                          
    \begin{tabular}{c c c c c c}        
    \hline\hline                 
    area & spectrum & band & 1st irr. & 2nd irr. & 3rd irr. \\    
    \hline                        
    b1 & 1 & 679 & -15 & 3 & 4 \\ 
    d1 & 1 & 664 & -1 & -3 & 7 \\ 
    d1 & 2 & 660 & 3 & 1 & 3 \\ 
    d1 & 3 & 660 & 3 & 1 & 3 \\ 
    d1 & 4 & 664 & -1 & 1 & 3 \\ 
    d1 & 5 & 664 & -1 & 1 & 3 \\ 
    a4 & 1 & 668 & -8 & -3 & 7 \\ 
    a5 & 1 & 668 & -8 & 0 & 4 \\ 
    a5 & 2 & 668 & -1 & 0 & -3 \\ 
      & average & 666 & -3.22 & 0.11 & 3.44 \\ 
      & min & 660 & -15 & -3 & -3 \\ 
      & max & 679 & 3 & 3 & 7 \\ 
      & median & 664 & -1 & 1 & 3 \\ 
    \hline                                   
    \end{tabular}
    \end{table}

      \begin{table*}[!hbt]
    \caption{Number of spectra with identified minerals according to major bands before the first and after the three irradiations.}             
    \label{tab:8}      
    \centering                          
    \begin{tabular}{c c c c c c c c c c c}
    \hline\hline
    mineral & Band & before irr. & ratio(\%) & 1st irr. & ratio (\%) & 2nd irr. & ratio(\%) & 3rd irr. & ratio (\%) & ratio range \\ 
    \hline
   no. of measurements & & 46 & & 26 & & 37 & & 49 & &  \\ 
    olivine & 849.2 & 6 & 13 & 2 & 8 & 15 & 41 & 6 & 12 & 8-13 \\ 
    olivine & 887.7 & 21 & 46 & 10 & 38 & 10 & 27 & 6 & 12 & 12-46 \\ 
    feldspar & 1152.1 & 46 & 100 & 26 & 100 & 37 & 100 & 49 & 100 & 100 \\ 
    pyroxene & 1047.2 & 30 & 65 & 16 & 62 & 28 & 76 & 33 & 67 & 62-76 \\ 
    spinel & 666.1 & 38 & 83 & 18 & 69 & 35 & 95 & 41 & 84 & 69-95 \\ 
    \hline 
    \end{tabular}
    \end{table*}

     \begin{table*}[h!]
    \caption{Summary of the appearance and changes of minerals (main peaks) before and after the irradiations.}             
    \label{tab:9}      
    \centering                          
    \begin{tabular}{c c c c c}
    \hline\hline
   mineral & appear later & disappeared later & cyclic appearance & occur always \\ 
   \hline
spinel & 17 (11\%) & 5 (3\%) & 17 (11\%) & 9 (6\%) \\ 
pyroxene & 11 (7\%) & 15 (9.5\%) & 15 (9.5\%) & 16 (10\%) \\ 
feldspar & 10 (6\%) & 8 (5\%) & 18 (11\%) & 22 (14\%) \\ 
olivine & 16 (10\%) & 11 (7\%) & 7 (4\%) & 5 (3\%) \\ 
    \hline 
    \end{tabular}
    \tablefoot{(Before irradiation: 46 identified mineral spectra, after first irradiation: 26 spectra, after second irradiation: 37 spectra, after third irradiation: 49 spectra.) In the above cases, all of the band observations were successful both after and before the irradiations, while the cyclic band appearance indicates that the given band emerged earlier, later disappeared, and finally emerged again.}
    \end{table*}

       \begin{table*}[h!]
    \caption{Average peak position shift summary by major bands of minerals.}             
    \label{tab:10}      
    \centering                          
    \begin{tabular}{c c c c c c c c c}
    \hline\hline
   mineral & band & 1st irr. & deviation & 2nd irr. & deviation & 3rd irr. &  deviation & avg. deviation after irr. \\ 
   \hline
olivine & 849.2 & -11.33 & 6.5 & 2 & 12.8 & 1.8 & 5.7 & 8.3 \\ 
olivine & 887.8 & 4.4 & 3.6 & -4 & 3.6 & 0.6 & 4.8 & 4 \\ 
feldspar & 1152.1 & -1.58 & 20.4 & 2.16 & 19.9 & -1.4 & 15.7 & 18.7 \\ 
pyroxene & 1047.2 & -7.8 & 14.9 & 7 & 19.6 & 1.4 & 5.6 & 13.4 \\ 
spinel & 666.1 & -3.22 & 5.9 & 0.11 & 1.9 & 3.4 & 2.9 & 3.6 \\  
    \hline 
    \end{tabular}
    \tablefoot{Average peak position shifts, where minerals occur after all of the irradiations (spinel, pyroxene, feldspar, but not olivine because of less data). The standard deviation of peak shift is the difference from the average values of peak shift position.}
    \end{table*}

     \begin{table*}[h!]
    \caption{Average FWHM values of major mineral bands with standard deviations.}             
    \label{tab:11}      
    \centering                          
    \begin{tabular}{c c c c c c c c c c c}
    \hline\hline
   mineral & band & 1st irr. & deviation & 2nd irr. & deviation & 3rd irr. &  deviation & avg. deviation after irr. \\
   \hline
olivine & 849.3 & 9.8 & 4.4 & 17.5 & 4.5 & 6.8 & 1 & 4.3 & 1.03 & 2.2 \\ 
olivine & 887.6 & 12 & 3.1 & 8.4 & 2.2 & 6.5 & 2.5 & 12 & 13.9 & 6.2 \\ 
feldspar & 1149.8 & 76.4 & 31.4 & 86.3 & 25.2 & 84.8 & 30.4 & 99.8 & 19.8 & 25.1 \\ 
pyroxene & 1048.4 & 29.2 & 14 & 22.1 & 11.2 & 22.3 & 20.5 & 21.3 & 9.03 & 13.6 \\ 
spinel & 668.5 & 65.8 & 40.7 & 51.4 & 46.3 & 65.6 & 41.9 & 76.6 & 33.09 & 40.4 \\  
    \hline 
    \end{tabular}
    \tablefoot{At the end of the table the average deviation after the irradiations was calculated. This table contains these measurements, where the given bands appeared before and after all irradiation tests.}
    \end{table*}

\end{appendix}

%
%

\end{document}